\documentstyle[12pt,epsfig,a4p,amssymb,amsbsy,hhline,cite,pstricks]{article}
\newcommand{\mr}{\mathrm}
\newcommand{\ee}{e$^+$e$^-$}
\newcommand{\eeto}{e$^+$e$^- \rightarrow$}
\newcommand{\qq}{${\mr q\bar{q}}$}
\newcommand{\qqg}{${\mr q\bar{q}}$g}
\newcommand{\qqgb}{${\mr q\bar{q}}$(g)}
\newcommand{\xe}{$x_{\mr E}$}
\newcommand{\Qj}{$Q_{\mr jet}$}
\newcommand{\ff}{fragmentation function}
\newcommand{\ffs}{fragmentation functions}
\newcommand{\CLEO}{CLEO Collaboration}
\newcommand{\OP}{OPAL Collaboration}
\newcommand{\DE}{DELPHI Collaboration}
\newcommand{\AL}{ALEPH Collaboration}
\newcommand{\SLD}{SLD Collaboration}
\newcommand{\PRL}{Phys. Rev. Lett.}
\newcommand{\PL}{Phys. Lett.}
\newcommand{\PR}{Phys. Rep.}
\newcommand{\PRV}{Phys. Rev.}
\newcommand{\EPJ}{Eur. Phys. J.}
\newcommand{\ZP}{Z. Phys.}
\newcommand{\NP}{Nucl. Phys.}
\newcommand{\NIM}{Nucl. Instr. and Meth.}
\newcommand{\CPC}{Comp. Phys. Commun.}
\newcommand{\al}{{\it et al.}}

\newcommand{\n}{\normalsize}
\newcommand{\FFl}{\n $\frac{1}{N_{\mr jet}^{\mr udsc}}\frac{{\mr d}N_{\mr ch}}
{{\mr d}x_{\mr E}}$}
\newcommand{\FFb}{\n $\frac{1}{N_{\mr jet}^{\mr b}}\frac{{\mr d}N_{\mr ch}}
{{\mr d}x_{\mr E}}$}
\newcommand{\FFg}{\n $\frac{1}{N_{\mr jet}^{\mr g}}\frac{{\mr d}N_{\mr ch}}
{{\mr d}x_{\mr E}}$}
\newcommand{\FFi}{\n $\frac{1}{N_{\mr jet}}\frac{{\mr d}N_{\mr ch}}{{\mr d}
x_{\mr E}}$}
\newcommand{\x}{\n $x_{\mr E}$}
\newcommand{\Q}{\n $Q_{\mr jet}$ [GeV]}
\newcommand{\sa}{\n scale \small [GeV]}
\newcommand{\p}{$\!\!\!\!\pm$}

\begin{document}
\renewcommand{\Huge}{\huge}
\thispagestyle{empty}
\addtocounter{page}{-1}

\begin{titlepage}

\begin{center}
{\large EUROPEAN ORGANIZATION FOR NUCLEAR RESEARCH} \\
\vspace*{0.7cm}

\begin{flushright}
CERN-PH-EP/2004-011\\
24 March 2004
\end{flushright}
\vspace*{0.7cm}

{\huge \bf Scaling violations of quark and gluon \\
jet \ffs\ \\
in \ee\ annihilations \\[4mm]
at $\mathbf \sqrt s =$ 91.2 and 183--209~GeV}
\vspace*{1.5cm}\\
{\LARGE The OPAL Collaboration}
\end{center}

\vspace*{0.6cm}
\begin{center}
{\large \bf Abstract}
\begin{quotation}

\noindent
Flavour inclusive, udsc and b \ffs\ in unbiased jets, and
flavour inclusive, udsc, b and gluon \ffs\ in biased jets 
are measured in \ee\ annihilations from data collected at 
centre-of-mass energies of 91.2, and 183--209~GeV with the OPAL 
detector at LEP. The unbiased jets are defined by hemispheres of
inclusive hadronic events, while the biased jet measurements are 
based on three-jet events selected with jet algorithms.
Several methods are employed to extract the \ffs\ over a wide range 
of scales. Possible biases are studied in the results obtained. The \ffs\ 
are compared to results from lower energy \ee\ experiments and with 
earlier LEP measurements and are found to be consistent. 
Scaling violations are observed and are found to be stronger for the 
\ffs\ of gluon jets than for those of quarks. 
The measured \ffs\ are compared to three recent theoretical next-to-leading 
order calculations and to the predictions of three Monte Carlo 
event generators. While the Monte Carlo models are in good agreement
with the data, the theoretical predictions fail to describe the full
set of results, in particular the b and gluon jet measurements.

\end{quotation}
\vspace*{1.1cm}
(\large \it to be submitted to European Physical Journal C)
\end{center}

\end{titlepage}

\begin{center}{\Large        The OPAL Collaboration
}\end{center}\bigskip
\begin{center}{
G.\thinspace Abbiendi$^{  2}$,
C.\thinspace Ainsley$^{  5}$,
P.F.\thinspace {\AA}kesson$^{  3,  y}$,
G.\thinspace Alexander$^{ 22}$,
J.\thinspace Allison$^{ 16}$,
P.\thinspace Amaral$^{  9}$, 
G.\thinspace Anagnostou$^{  1}$,
K.J.\thinspace Anderson$^{  9}$,
S.\thinspace Asai$^{ 23}$,
D.\thinspace Axen$^{ 27}$
% ...
}\end{center}\bigskip
\bigskip
$^{  1}$School of Physics and Astronomy, University of Birmingham,
Birmingham B15 2TT, UK
\newline
$^{  2}$Dipartimento di Fisica dell' Universit\`a di Bologna and INFN,
I-40126 Bologna, Italy
\newline
$^{  3}$Physikalisches Institut, Universit\"at Bonn,
D-53115 Bonn, Germany
\newline
% ...
$^{  5}$Cavendish Laboratory, Cambridge CB3 0HE, UK
\newline
% ...
$^{  9}$Enrico Fermi Institute and Department of Physics,
University of Chicago, Chicago IL 60637, USA
\newline
% ...
$^{ 16}$Department of Physics, Schuster Laboratory, The University,
Manchester M13 9PL, UK
\newline
% ...
$^{ 22}$Department of Physics and Astronomy, Tel Aviv University,
Tel Aviv 69978, Israel
\newline
$^{ 23}$International Centre for Elementary Particle Physics and
Department of Physics, University of Tokyo, Tokyo 113-0033, and
Kobe University, Kobe 657-8501, Japan
\newline
% ...
$^{ 27}$University of British Columbia, Department of Physics,
Vancouver BC V6T 1Z1, Canada
\newline
% ...
$^{  y}$ now at CERN
\newline
% ...

\section{Introduction}
Hadron production in high energy collisions can be described by parton showers
(successive gluon emissions and  splittings), followed by the formation of 
hadrons which cannot be described perturbatively. Gluon emission, the dominant
process in parton showers, is proportional to the colour factor associated
with the coupling of the emitted gluon to the emitter. These colour factors 
are $C_A=3$ when the emitter is a gluon and $C_F=4/3$ when it is a quark. 
Consequently, the multiplicity of soft gluons from a gluon source is 
(asymptotically) 9/4 times higher than from a quark source \cite{Brodsky}.
The inequality between $C_A$ and $C_F$ plays a key role in the explanation of 
the observed differences between quark and gluon jets: compared to quark jets, 
gluon jets are observed to have larger widths \cite{qgdif}, higher 
multiplicities \cite{qgdif,subjet}, softer \ffs\ \cite{qgdif,gincl3,Master}, 
and stronger scaling violations of the \ffs\ \cite{Master}. \\

The \ff, $D_a^h(x,Q^2)$, is defined as the probability that parton $a$, which 
is produced at short distance, of order $1/Q$, fragments into hadron, $h$, 
carrying fraction $x$ of the momentum of $a$. In this study, the momentum 
fraction is defined as \xe\ $=E_h/E_{\mr jet}$, where $E_h$ is the energy of 
the hadron $h$ and $E_{\mr jet}$ is the energy of the jet to which it is 
assigned. The relative softness of the gluon jet \ff\ is explained in the low 
\xe\ region by the higher multiplicity of soft gluons radiated,
and in the high \xe\ region by the fact 
that the gluon cannot be present as a valence parton inside a produced hadron 
(first a splitting ${\mr g\rightarrow q\bar{q}}$ has to occur).
The stronger scaling violation is due to the fact that the scale dependence of
the gluon jet \ff\ is dominated by the splitting function $P_{\mr g\rightarrow 
gg} \sim C_A$, while that of the quark jet is dominated by the splitting 
function $P_{\mr q\rightarrow qg} \sim C_F$.\\
 
Jets in \ee\ annihilations are commonly defined using a jet finding algorithm,
which is a mathematical prescription for dividing an event into parts
associated with individual quarks and gluons. For example, quark and gluon 
jets are often defined by applying a jet finder to select three-jet \qqg\ 
events. Some of the most common algorithms are the Durham \cite{Durham} and 
cone \cite{cone} jet finders. Different jet finders result in different 
assignments of particles to jets: thus jets defined using a jet finding
algorithm are called {\it biased}. In contrast, quark and gluon jets used in 
theoretical calculations are usually based on inclusive samples of back-to-back
\qq\ and gg final states rather than three-jet events. A hemisphere of a \qq\ 
event is defined as a quark jet and similarly, a gluon jet is defined by a 
hemisphere in a gg final state. The hemisphere 
definition yields a so-called {\it unbiased} jet because the jet properties do
not depend on the choice of a jet finder. Measurements of unbiased quark jets
have been performed at many scales since such jets correspond to hemispheres 
of inclusive \eeto\ hadrons events \cite{loweren,B-decay,flavourff}. Direct 
measurements of unbiased gluon jets are so far available only from the CLEO 
\cite{CLEO} and OPAL \cite{gincl3,gincl12} experiments, however. At CLEO, jets 
originating from radiative $\Upsilon$ decays have energies of only about 
5~GeV, which limits their usefulness for jet studies. 
In \cite{gincl3,gincl12}, unbiased gluon jets were selected using rare events 
of the type \eeto\ ${\mr q\bar{q}g_{incl}}$, in which the object 
${\mr g_{incl}}$, taken to be the gluon jet, is defined by all particles 
observed in the hemisphere opposite to that containing two b-tagged quark 
jets. Due to the low probability of such a topology, this method of obtaining 
unbiased gluon jets is only viable for very high statistics data samples. 
Recently, the OPAL experiment has measured properties of unbiased gluon jets 
indirectly. In \cite{indgl1}, recent theoretical expressions to account for 
biases from event selection were used to extract gluon jet properties over a 
range of jet energies from about 11 to 30 GeV. In \cite{indgl2}, the first 
experimental results based on the so-called jet boost algorithm, a technique 
to select unbiased gluon jets in \ee\ annihilations, were presented for jet 
energies from 5 to 18 GeV.\\

Scaling violations of quark and gluon jet \ffs\ from three-jet events produced 
in \ee\ collisions at a center-of-mass system (c.m.s.) energy of $\sqrt s =$ 
91.2~GeV, based on the $k_T$ jet algorithms Durham \cite{Durham} and 
Cambridge \cite{Cambridge}, were reported in \cite{Master}.
These scaling violations were found to be consistent with the expectations from
the Dokshitzer-Gribov-Lipatov-Altarelli-Parisi (DGLAP) evolution equations
\cite{DGLAP}. In our study, we present measurements of quark 
and gluon jet \ffs\ at $\sqrt s =$ 91.2~GeV and $\sqrt s =$ 183--209~GeV. 
The data were collected with the OPAL detector at the LEP \ee\ collider at 
CERN. We measured seven types of \ffs: the udsc, b, gluon and flavour 
inclusive \ffs\ in biased jets, and the udsc, b, and flavour inclusive \ffs\ 
in unbiased jets. While the two types of flavour inclusive \ffs\ have been 
measured many times, data on the other types of \ffs\ are still rather 
scarce. \\

The paper is organised as follows. In Section~\ref{detector}, a brief 
description of the OPAL detector is given. The samples of data and simulated
events used in the analysis are described in Section~\ref{samples}. In 
Section~\ref{evsel}, the event and jet selections are discussed. The analysis 
procedure, including the methods used to evaluate systematic uncertainties, is
presented in Section~\ref{procedure}. Section~\ref{bias} deals with a Monte 
Carlo (MC) study of the biases introduced by our jet finding procedure. 
Next-to-leading order (NLO) calculations \cite{KKP,Kretzer,BFGW} for \ffs\ 
are described in Section~\ref{NLO}. 
In Section~\ref{results}, we present a comparison of our data to other 
measurements, to MC predictions, and to the NLO calculations. A summary and 
conclusions are given in Section~\ref{conclusions}.

\section{The OPAL detector}\label{detector} 
The OPAL detector is described in detail elsewhere \cite{OPAL}. The tracking 
system consists of a silicon microvertex detector, an inner vertex chamber,
a large volume jet chamber and specialized chambers at the outer radius of 
the jet chamber which improve the measurements in the $z$-direction\footnote{OPAL uses a right-handed coordinate system defined with positive $z$ along the 
electron beam direction and with positive $x$ pointing towards the centre of 
the LEP ring. 
The polar angle $\theta$ is defined relative to the $+z$ axis and the azimuthal
angle $\phi$ relative to the $+x$ axis.}. The tracking system covers the 
region $|\cos\theta|<0.98$ and is enclosed by a solenoidal magnet with an 
axial field of 0.435 T. Electromagnetic energy is measured by a lead-glass 
calorimeter located outside the magnet coil, which covers $|\cos\theta|<0.98$.

\section{Data and Monte Carlo samples} \label{samples}
The present analysis is based on two data samples which will be referred to 
as the LEP1 and LEP2 samples. The LEP1 data sample contains hadronic Z 
decay events collected with the OPAL detector between 1993 and 1995 at 
c.m.s.\ energies within 250 MeV of the Z peak. The LEP2 data sample contains 
hadronic events collected with the OPAL detector in the period 1997--2000
at c.m.s.\ energies in the range 183--209 GeV. All the data were taken with 
full readout of the $r$-$\phi$ and $z$ coordinates of the silicon 
microvertex detector which is essential for precise measurements of primary 
and secondary vertices. The total integrated luminosity in the LEP1 data is
130~pb$^{-1}$, while the LEP2 data sample corresponds to a luminosity of 
690~pb$^{-1}$. \\

In this study, we work with three types of MC event samples. The detector 
level samples include full simulation of the detector response \cite{simul}, 
the initial-state photon radiation (ISR) and background processes, and 
contain only those events which pass the same selection cuts as applied to 
the data. The hadron level samples do not include ISR or detector simulation 
and allow all particles with lifetimes shorter than 3 $\times 10^{-10}$ s to 
decay. The parton level samples are formed by final-state partons, i.e. quarks 
and gluons present at the end of the perturbative shower, and do not include
ISR.\\

Signal MC events for the LEP1 data, of the form \eeto\ Z $\rightarrow$ \qqgb, 
are generated using the JETSET~7.4 \cite{jetset7.4} and HERWIG~6.2 
\cite{herwig6.2} programs with the parameter settings tuned on LEP1 OPAL data
described in \cite{set_js7.4} and \cite{set_hw6.2}, respectively. For LEP2 
data, the signal \eeto\ Z$^*/\gamma^*\rightarrow$ \qqgb\ events are simulated 
using PYTHIA~6.125 \cite{jetset7.4, pythia6.125} and HERWIG~6.2. For events of 
this type, PYTHIA is the same as JETSET except for the inclusion of ISR 
processes. 
The same parameter settings are used for the LEP2 PYTHIA and HERWIG samples 
as are used for the LEP1 JETSET and HERWIG samples, respectively. In the 
JETSET and PYTHIA event generators, the string fragmentation model is 
implemented, while HERWIG uses the cluster fragmentation model. The initial- 
and final-state photon radiation for the LEP2 MC samples are performed by 
interfacing the KK2F program \cite{kk2f} to the main generator programs.
In addition to PYTHIA and HERWIG we also use the ARIADNE~4.08 \cite{ariadne}
event generator to compare with the final results. For hadronization, the 
generator is interfaced to the JETSET~7.4 program. The parameter settings used
for ARIADNE are documented in \cite{gincl3,aleph-ar}. \\  

To estimate the background in the LEP2 data, we generate events of the type
\ee $\rightarrow$ 4 fermions. These events, in particular those with four
quarks in the final state, constitute the major background in this analysis.
The 4-fermion events are generated using the GRC4F 2.1 \cite{grc4f} MC event 
program. The final states are produced via s-channel or t-channel diagrams 
and include W$^+$W$^-$ and ZZ events. This generator is interfaced 
to PYTHIA using the same parameter set for the fragmentation and decays as 
used for the signal events.\\

The signal as well as the background MC event samples for the LEP2 period 
are generated at c.m.s.\ energies of 183, 189, 192, 196, 200, 202, 204, 
205, 206, 207 and 208 GeV reflecting the energy distribution in the collected
data samples. 

\section{Event and jet selection}\label{evsel}

\subsection{Selection of hadronic Z and Z$\mathbf ^*/\gamma^*$ events}\label{evesel}
The procedures for identifying hadronic events are discussed in \cite{hadr}.
The selection of the inclusive hadronic event sample in the LEP1 data is 
based on tracks and electromagnetic clusters. Tracks are required to have at 
least 40 measured
points (of 159 possible) in the jet chamber, to have a momentum greater 
than 0.15~GeV/$c$, to lie in the region $|\cos\theta|$$\,<\,$0.94, to have a
distance of the point of closest approach to the collision point in the 
$r$-$\phi$ plane, $d_0\le$ 5~cm, and along the $z$ axis, $z_0\le$ 25~cm.
Clusters are required to be spread 
over at least two lead glass blocks and to have an energy greater than 
0.10~GeV if they are in the barrel section of the detector 
($|\cos\theta|$$\,<\,$0.82) or greater than 0.20~GeV if they are in the 
endcap section (0.82$\,<\,$$|\cos\theta|$$\,<\,$0.98). 
A matching algorithm is employed to reduce double counting of energy in cases 
where tracks point towards electromagnetic clusters. Specifically, the expected
calorimeter energy of the associated tracks is subtracted from the cluster
energy. If the energy of a cluster is smaller than that expected for the 
associated tracks, the cluster is not used. Each accepted track and cluster 
is considered to be a particle. Tracks are assigned the pion mass. Clusters 
are assigned zero mass since they originate mostly from photons.\\

To eliminate residual backgrounds, the number of accepted tracks in each event 
is required to be at least five. To reject events in which a significant 
number of particles is lost near the beam line direction, the thrust 
axis of the event, calculated using the particles, is required to satisfy 
$|\cos (\theta_{\rm thrust})|$$\,<\,$0.90, where $\theta_{\rm thrust}$ is the 
angle between the thrust and beam axes.
The two-photon background (events of the type $\gamma\gamma\rightarrow$ \qq)
is reduced by imposing the conditions 
$E_{\mr vis}/\sqrt s > 0.1$ and $|p_{\rm bal}|<0.6$, where $E_{\mr vis}$ is 
the total visible energy (i.e. the sum of detected particle energies) and 
$p_{\rm bal}$ is the momentum sum in the $z$ direction, normalized by 
$E_{\mr vis}$. The residual background in the LEP1 data sample from all 
sources is estimated to be less than~1\% \cite{hadr}. The number of 
inclusive hadronic events is $2\, 387\, 227$ (see the first row in Table 
\ref{LEP12stat}), with the selection efficiency estimated to be 96\%. \\

At c.m.s.\ energies above the Z resonance, several new sources of 
background exist. To select hadronic events in the LEP2 data, the same 
procedure as described for the LEP1 data is used and in addition, we apply the
procedure described in \cite{OP133,OP161,OP172,Joost} to reduce the 
background as summarized below. \\
 
The majority of hadronic events at LEP2 are radiative events in which 
initial-state 
radiation reduces the original c.m.s.\ energy of the hadronic system. To reject
such ISR events, we determine the effective c.m.s.\ energy of the hadronic 
system, $\sqrt {s^\prime}$, following the procedure described in \cite{Joost} 
which takes possible multiple photon radiation into account. We require  
$\sqrt s - \sqrt {s^\prime} <$ 10 (20) GeV to select inclusive hadronic 
events for the hemisphere (three-jet) analysis described below. We refer to 
this procedure as the ``invariant mass'' selection. 
For systematic studies, we apply an alternative method based on 
combining cuts on the visible energy and missing momentum of the event and on 
the energy of an isolated photon candidate \cite{OP133}. This procedure is 
referred to as the ``energy balance'' selection. Simulated hadronic 
Z$^*/\gamma^*$ events are defined to be radiative if $\sqrt{s_{\mr true}^{\prime}}<\sqrt s-1$~GeV, where $\sqrt{s_{\mr true}^{\prime}}$ is the true effective 
c.m.s.\ energy. The efficiency for selecting LEP2 non-radiative hadronic events
is 73\%.\\

The production of W$^+$W$^-$ and ZZ 
pairs with hadronic or semi-leptonic decays (4-fermion final states) is an
additional source of background. This background is reduced by applying a 
method described in \cite{Joost}: first each event is forced into a four-jet 
configuration using the Durham jet finder. In the LEP2 samples, the 4-momenta 
of all measured particles are boosted into the rest frame of the hadronic 
system with the effective c.m.s. energy, $\sqrt {s^\prime}$, and are then used
to find jets. Then an event weight $W_{\mr QCD}$ 
is defined based on calculated QCD matrix elements for the process \eeto\ 
${\mr q\bar{q}q\bar{q}}$ or ${\mr q\bar{q}gg}$, with the four parton final 
state corresponding to the obtained four-jet kinematics \cite{terano}. The QCD
matrix elements are calculated using the EVENT2 program \cite{event2}. 
A good separation between the Z$^*/\gamma^*$ and W$^+$W$^-$ or ZZ 
pair events is achieved by requiring $W_{\mr QCD}\ge-0.5$.\\

The remaining background from \eeto\ ${\mr \tau^+\tau^-}$ and two-photon 
events is estimated to be about 0.2\% \cite{Joost} and is neglected. 
The remaining 4-fermion background is subtracted from the data bin-by-bin. 
The number of the inclusive hadronic events in the LEP2 data sample for the 
hemisphere (three-jet) analysis is $10\, 866$ ($12\, 653$) with 11\% (14\%) 
4-fermion background (see the first row in Table \ref{LEP12stat}).  

\subsection{Jet selection}\label{jetsel}
As explained in the introduction, we employ two definitions of jets. In the 
inclusive hadronic event samples we use the unbiased jet definition where the 
jets are defined by particles in hemispheres of the \qq\ system. 
In the three-jet samples, we apply a jet algorithm and thus work with biased 
jets. Three jet algorithms are used: the Durham \cite{Durham}, Cambridge 
\cite{Cambridge} and cone \cite{cone} algorithms. Relatively large differences
in the techniques used by the $k_T$ and cone jet finders ensure that the jet 
finder dependence of the results is estimated conservatively. The jet 
algorithm is forced to resolve three jets in each event. The jet energies and 
momenta are then recalculated by imposing overall energy-momentum conservation 
with planar massless kinematics, using the jet directions found by the jet 
algorithm. The jet energies are given by the relation:
\begin{equation}
 E_i=\frac{\sqrt s \cdot \sin \theta_{j,k}}{\sin \theta_{i,j}+
\sin \theta_{j,k}+\sin \theta_{k,i}} 
\label{ejcal}
\end{equation}
where $\theta_{i,j}$ is the angle between jets $i$ and $j$ and $k$ corresponds
to the remaining jet. We note that for the LEP2 detector level jets, the 
effective c.m.s.\ energy, $\sqrt {s^\prime}$, is used in the above formula.
Cuts, given in Table~\ref{Jetcuts}, are chosen to ensure
that the jets are well contained within the sensitive part of the detector, 
well separated from each other and that the event is planar. The numbers of 
LEP1 and LEP2 events passing these selection criteria are shown in the second
row of Table~\ref{LEP12stat}. The efficiency for selecting non-radiative 
three-jet LEP2 events is 68\%.\\

All three jet algorithms yield very similar jet angular and energy resolutions,
with the Durham algorithm being slightly better than the other two. Therefore,
the Durham algorithm is used as the reference, with the cone and Cambridge jet
finders used for systematic studies. The jet energy resolution, defined as
$(E_{\mr jet}^{\mr detector}-E_{\mr jet}^{\mr parton})/E_{\mr jet}^
{\mr parton}$, is found to range from 2\% for the most energetic jet to 
11\% for the least energetic jet. The distribution of the angles between the 
detector and parton jet axes is found to have an RMS of 0.05 radians for the 
most energetic jet and 0.16 radians for the least energetic jet. See Section 
\ref{btag} for an explanation of how the detector and parton level jets are 
associated with each other.

\section{Analysis procedure}\label{procedure}
In the following, we describe the method we use to determine the quark and 
gluon jet \ffs. The measured \ff\ is defined here as the total number of 
charged particles, $N_p$, in bins of \xe\ and scale $Q$ normalized to the 
number of jets, $N_{\mr jet}(Q)$, in the bin of $Q$:
\begin{equation}
\frac{1}{N_{\mr jet}(Q)}\frac{{\mathrm d}N_p(x_{\mr E},Q)}
{{\mathrm d}x_{\mr E}}
\label{ffdef}
\end{equation}
where \xe\ is defined in the Introduction.

\subsection{Jet scale $\mathbf Q_{\mr \mathbf jet}$}
\begin{figure}[tb] \centering
\epsfig{file=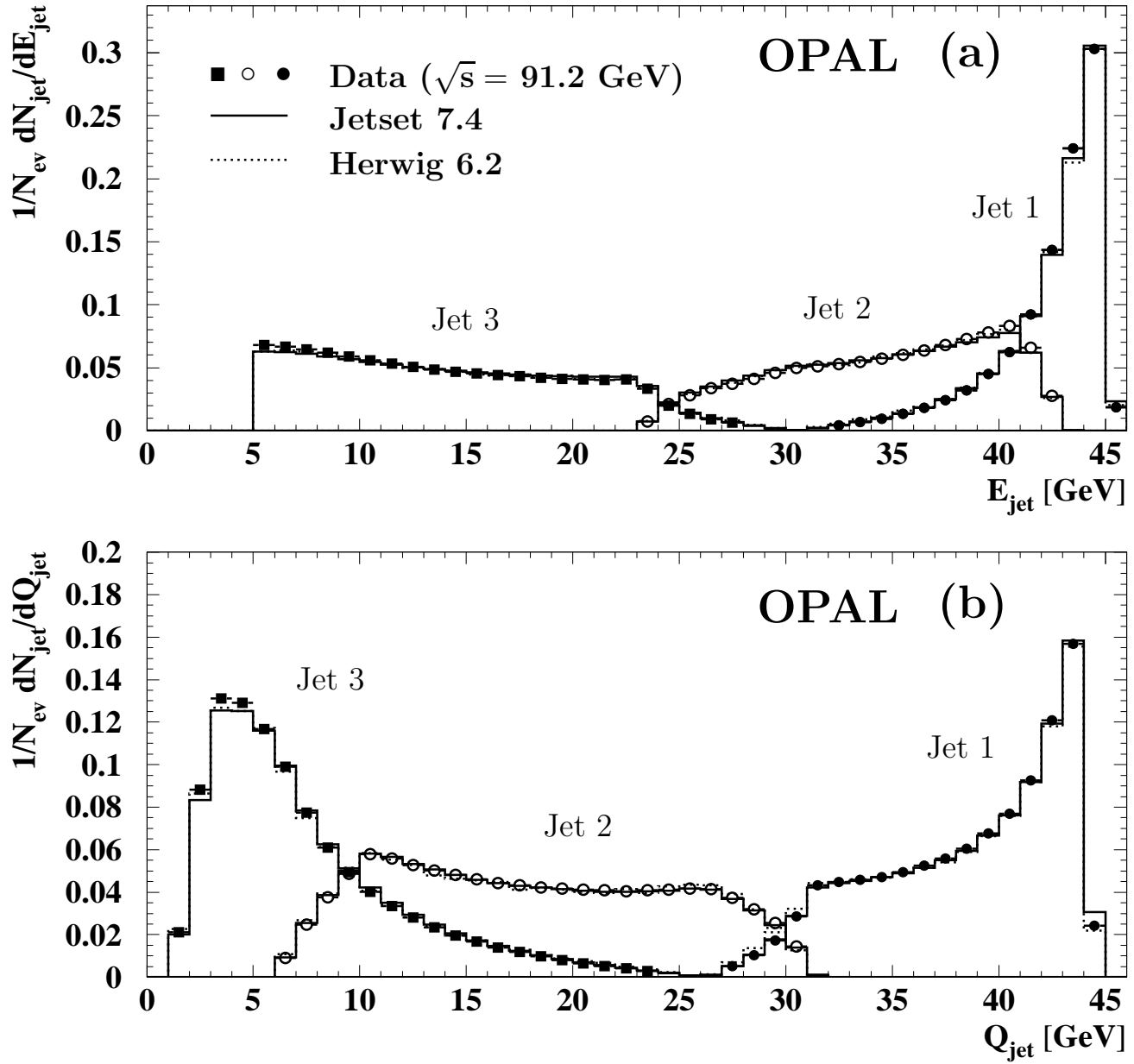,bbllx=40pt,bblly=150pt,bburx=530pt,%
bbury=671pt,width=\textwidth}
\caption{(a) Jet energy and (b) jet scale \Qj\ distributions
for the selected three-jet events in the LEP1 sample. The solid histograms 
represent the JETSET 7.4 and the dotted histograms the HERWIG 6.2 predictions 
using the Durham jet algorithm. The data are shown with statistical 
uncertainties only.}
\label{eqjet1}
\end{figure}

\begin{figure}[tb] \centering
\vspace*{-0.5cm}
\epsfig{file=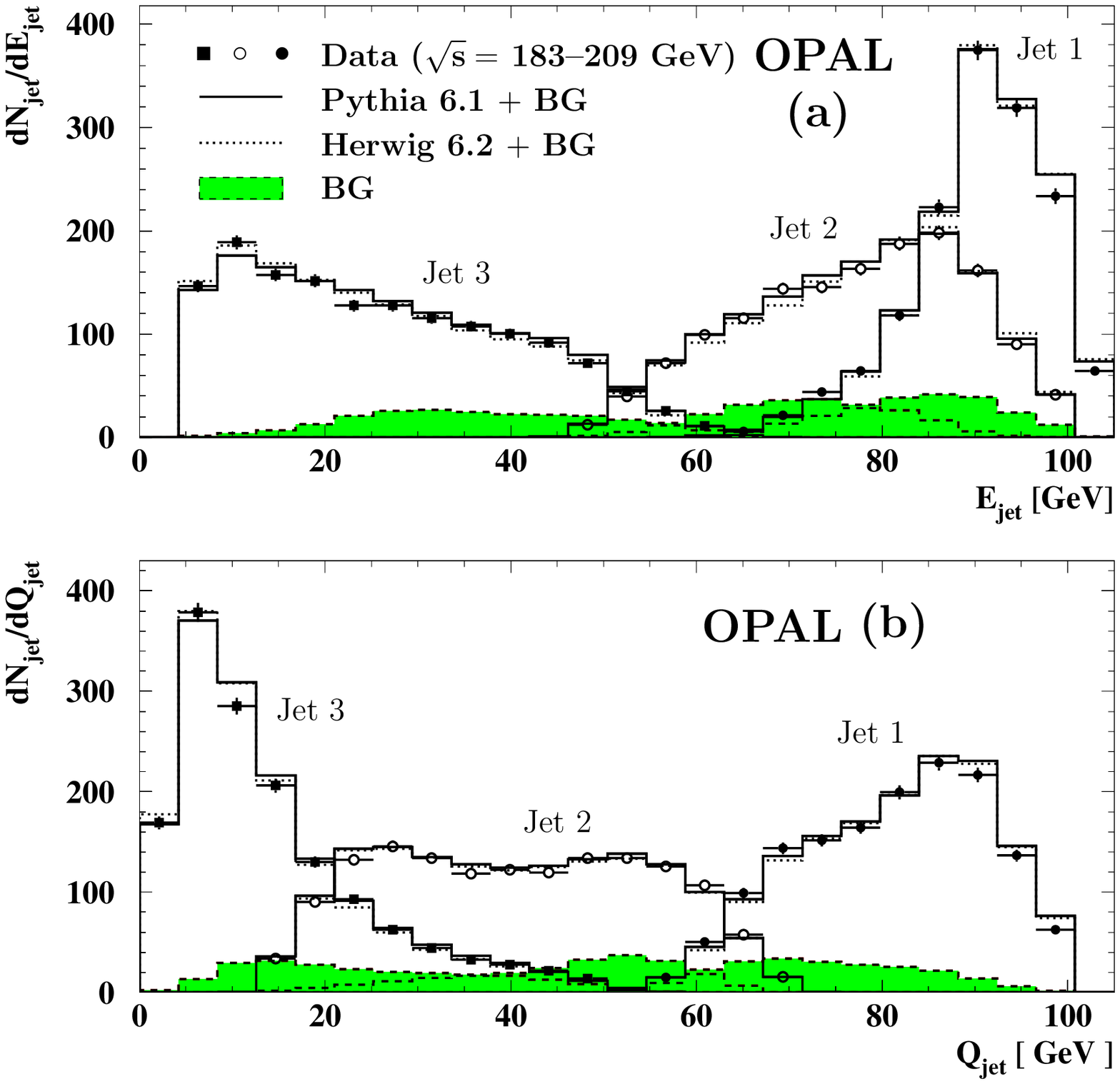,bbllx=37pt,bblly=205pt,bburx=530pt,%
bbury=677pt,width=\textwidth}
\vspace*{0.5cm}
\caption{(a) Jet energy and (b) jet scale \Qj\ distributions
for the selected three-jet events in the LEP2 sample. The solid histograms 
represent
the sum of the PYTHIA 6.125 and background (BG) predictions, the dotted 
histograms the sum of the HERWIG 6.2 and background predictions and the shaded
histograms the prediction of the model GRC4F for the background, all using the
Durham jet algorithm. The data are shown with statistical uncertainties only.}
\label{eqjet2}
\end{figure}

To measure the scale dependence, it is necessary to specify a scale 
relevant to the process under study. For inclusive hadronic events, the scale
is $\sqrt s$. For jets in three-jet events, neither $\sqrt s$ nor 
$E_{\mr jet}$ is considered to be an appropriate choice of the scale 
\cite{Qjet}. QCD coherence suggests \cite{QCDBas} that the event topology 
(i.e. the positions of the partons with respect to each other) should also be 
taken into account. In studies of quark and gluon jet characteristics 
\cite{Master,Qjet,Madjid,Dmult3} the transverse momentum-like scale 
$Q_{\mr jet}$, of a jet with energy $E_{\mr jet}$ has been used:
\begin{equation}
Q_{\mr jet} = E_{\mr jet}\sin\frac{\vartheta}{2}, 
\label{Qjet}
\end{equation}
where $\vartheta$ is the angle between this jet and the closest other jet.
This scale roughly expresses a maximum allowed transverse momentum 
(or virtuality) of gluons radiated in the showering process with respect to 
the initial parton, whilst still being associated with the same jet.
This definition of scale is adopted for the present analysis. The jet energy 
and \Qj\ spectra are shown in Figs.~\ref{eqjet1} and \ref{eqjet2} for the 
three jets found by the Durham jet algorithm and ordered in energy, with jet~1
being the most energetic and jet~3 the least energetic jet. The data are seen 
to be well described by the JETSET and HERWIG models. A similar description
is also seen for the cone and Cambridge jet finders (not shown). 

\subsection{Quark and gluon jet identification}\label{identif}
There are several ways to identify quark and gluon jets. In this analysis, 
three methods are used: the b-tag and the energy-ordering methods to identify 
quark and gluon jets in biased three-jet events, and the hemisphere method to 
identify unbiased quark jets in inclusive hadronic events. In addition, b 
tagging is used to separate udsc and b quark jets from each other, both for 
the biased and unbiased jet samples. In contrast to the b-tag method, the 
energy-ordering 
method only allows flavour inclusive quark jets to be distinguished from gluon
jets. Note that the flavour composition of the primary quarks in \eeto\ \qq\ is
predicted by electroweak theory to vary with c.m.s.\ energy. Therefore, to 
perform a meaningful comparison of the biased jet data taken at $\sqrt s =$ 
91.2~GeV with the unbiased jet data measured at several c.m.s.\ energies, a 
special correction is applied in the construction of the flavour inclusive 
\ff\ from biased jets (see Section \ref{ficor}). \\

\subsubsection{b-tag method in three-jet events}\label{btag}
In the three-jet sample, the b-tagging technique is used to obtain samples 
enriched in udsc, b or gluon jets. The analysis utilizes an inclusive single 
jet tag method based on a neural network, as described in \cite{B-tag}. 
Any or all of the three jets may be used to extract the \ffs. 
Note that with our selection of three-jet events, the highest energy parton 
jet is predicted to be the gluon jet in 4.8\% of the events.\\

In the data and MC, three samples of jets are selected, each with different
fractions originating from udsc-quarks, b-quarks or gluons. We first look for 
jets with secondary vertices found in cones of radius $R=0.65$ radians from 
the jet axes. 
A jet is considered to be a b-tag jet if it contains a secondary vertex with 
neural network output value, VNN, greater than 0.8 for LEP1 events or 0.65 for 
LEP2 events. A jet with no secondary vertex, or a vertex with VNN$<$0.5 is 
considered to be an ``anti-tag'' jet. The b-tag and gluon jet samples are 
taken from events with one or two b-tag jets and at least one anti-tag jet. 
If one or two b-tag jets and one anti-tag jet are found, the b-tag jets 
enter the b-tag jet sample and the anti-tag jet enters the gluon jet sample.
If one b-tag jet and two anti-tag jets are found, the b-tag jet enters the 
b-tag jet sample, and the lower energy other jet is included in the gluon jet 
sample. 
The udsc jet sample is formed by all three jets in events with no b-tag jet or
with b-tag jets but no anti-tag jet (the contribution from the latter events
is negligible in practice).
Note that with this definition, the gluon jet is explicitly included in the 
udsc jet sample. The correction procedure to obtain a pure udsc jet sample 
with the gluon jet component removed is described below. \\

The purities of the different jet samples are evaluated by examining Monte 
Carlo events at the parton, hadron and detector levels. First, parton level 
jets are examined to determine whether they originate from a quark or a 
gluon. This determination is performed in two ways:
\begin{itemize}
\item {\bf Flavour assignment:}  
It is assumed that the highest momentum quark and antiquark with the correct
flavour for the event are the primary quark and antiquark. In events in which
different parton level jets contain the primary quark and antiquark, the 
remaining jet is assumed to arise from a gluon.

\item {\bf Non-flavour assignment:} A parton jet is identified as a quark 
(antiquark) jet if it contains an arbitrary number of \qq\ pairs 
and gluons plus one unpaired quark (antiquark). If such two parton jets are
found, the gluon jet is defined as that containing only \qq\ pairs
(if any) and gluons. 
\end{itemize}
A small fraction of events showing an ambiguous assignment of the primary
\qq\ pair and gluon to three parton level jets is excluded from the event 
samples. It amounts to 1.3\% for the flavour and 2.5\% for the non-flavour
assignment. To obtain the final results, the former method is used.\\

Detector and parton level jets are assigned to the hadron jet to which they 
are nearest in angle. For events in which more than one parton or detector 
level jets are assigned to the same hadron level jet (about 9\% of the 
events), the closest jet is chosen, while the more distant jet is assigned 
to the remaining hadron jet. The above procedure is referred to as the 
``matching'' procedure, and the hadron level jets associated with the parton 
level quark and anti-quark jets are defined to be {\it pure quark jets}, while
the remaining jet is a {\it pure gluon jet}.\\

\begin{figure}[tb] \centering
\epsfig{file=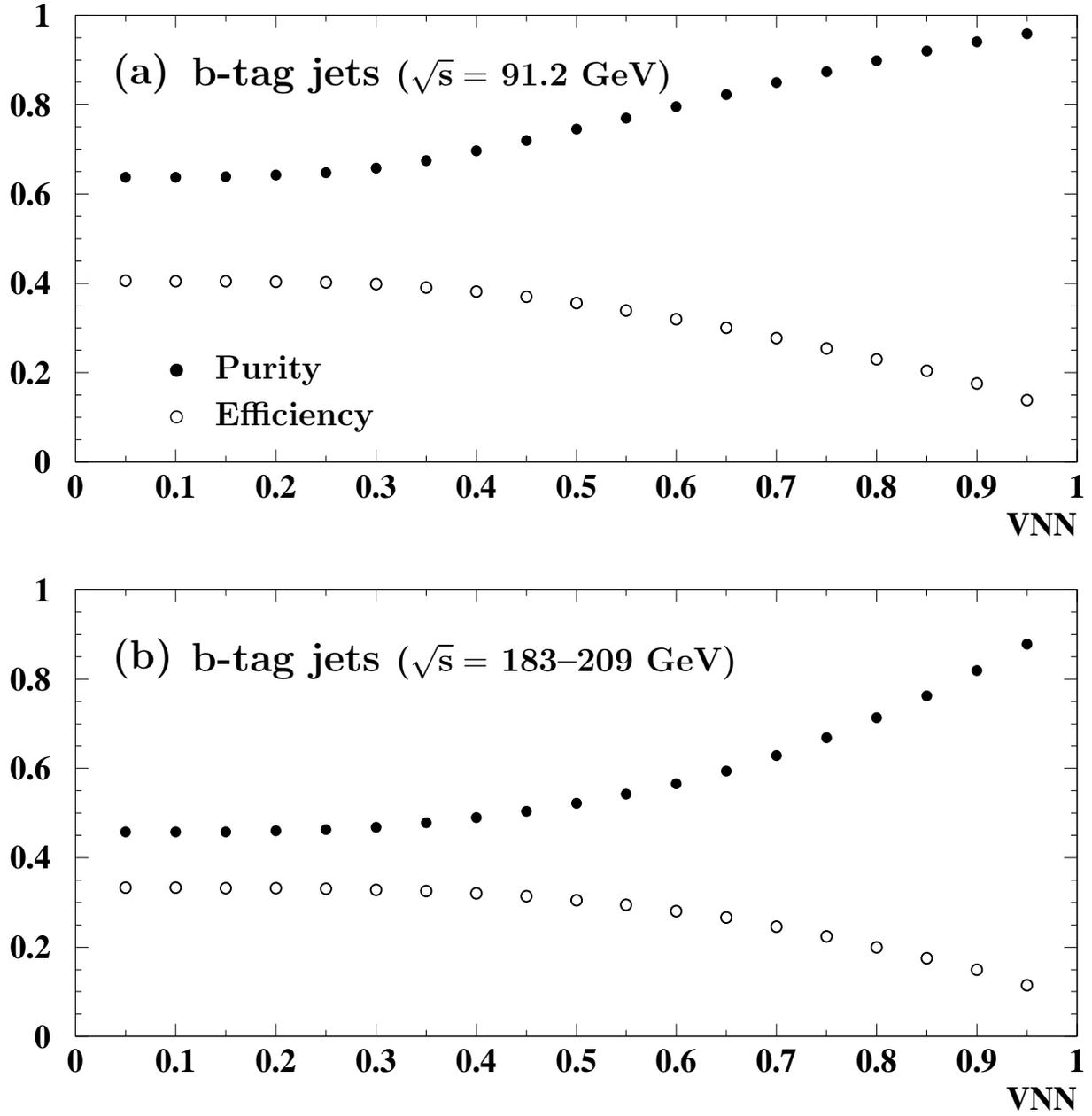,bbllx=68pt,bblly=217pt,bburx=522pt,%
bbury=683pt}
\vspace*{0.5cm}
\caption{The purity (solid circles) and efficiency (open circles) as a 
function of the neural network output VNN for (a) the LEP1 b-tag jet sample 
and (b) the LEP2 b-tag jet sample, obtained from the JETSET 7.4 for the LEP1 
events and PYTHIA 6.125 for the LEP2 events using the Durham jet algorithm.}
\label{pureff12}
\end{figure}

The purity and the efficiency of the LEP1 and LEP2 b-tag jet samples as a 
function of the VNN variable are shown in Fig.~\ref{pureff12}. The purity of 
the b-tag jet sample at the point VNN=X is defined as the fraction of pure b 
jets in the sample of b-tag jets with VNN$>$X. The efficiency of the b-tag jet
sample at the point VNN=X is defined as the fraction of the b-tag jets with 
VNN$>$X in the sample of all pure b jets. For VNN$>$0.8 applied in the LEP1 
samples, the purity of the b-tag jet sample is 90\% and the efficiency 23\%. 
The corresponding gluon jet purity and efficiency are 84\% and 40\%, 
respectively. The LEP2 samples are treated analogously to the LEP1 samples, 
except that we require VNN$>$0.65 because of low event statistics. The b 
(gluon) jet tagging efficiency is 27\% (45\%) and the purity 60\% (80\%). \\ 

To obtain a distribution of a variable $D$ (e.g. the \ff) of pure udsc 
(b, gluon) jets, $D_{\mr l(b,g)}^{\mr pure}$, one has to solve the following 
equation
\begin{equation}\stackrel{\mr \ \ \ \ \ uncor}{
\left( \begin{array}{l}D_{\mr l} \\
                       D_{\mr b} \\
                       D_{\mr g} \end{array} \right)\!(x_{\mr E},Q)} \ \ = \ \ 
\left( \begin{array}{lll}P_{\mr ll}~\:P_{\mr lb}~\:P_{\mr lg} \\
                         P_{\mr bl}~P_{\mr bb}~P_{\mr bg} \\
                         P_{\mr gl}~\,P_{\mr gb}~P_{\mr gg} \end{array} \right)\!(Q)
 \! \stackrel{\mr \ \ \ \ pure}{
\left( \begin{array}{l}D_{\mr l} \\
                       D_{\mr b} \\
                       D_{\mr g} \end{array} \right)\!(x_{\mr E},Q)}\\
\label{purmat}
\end{equation}
where $D_{\mr l(b,g)}^{^{\mr uncor}}$ stands for a distribution of the 
variable $D$ obtained from the sample of detector level udsc (b-tag, gluon) 
jets. The purity $P_{ij}$ denotes the probability that a jet from the jet 
sample {\it i} comes from a parton {\it j}. The indices {\it i, j} run over 
symbols l,b and g which stand for the u,d,s,c (``light'')-quark, b-quark and 
gluon. \\

\begin{figure}[tb] \centering
\vspace*{-0.8cm}
\epsfig{file=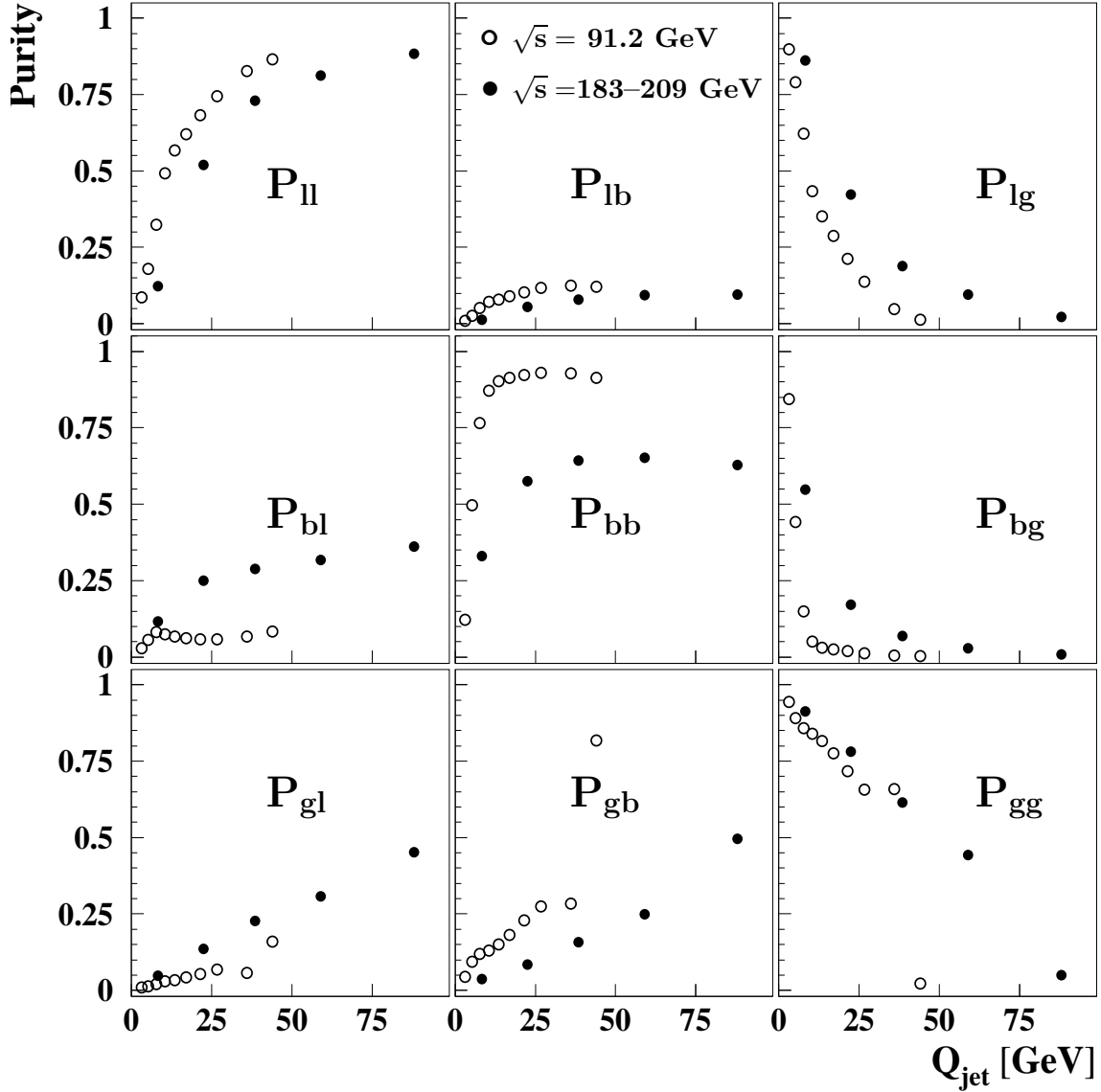,bbllx=82pt,bblly=219pt,bburx=510pt,%
bbury=641pt}
\vspace*{0.8cm}
\caption{The purity matrix as a function of \Qj\ scale for the LEP1 (open 
circles) and the LEP2 (solid circles) three-jet events, obtained from the 
JETSET 7.4 for the LEP1 events and PYTHIA 6.125 for the LEP2 events using 
the Durham jet algorithm. The flavour assignment is used for matching the 
primary outgoing partons to hadron jets (see text).}
\label{purmat1}
\end{figure}

In Fig.~\ref{purmat1} the LEP1 and LEP2 purity matrices as functions of \Qj\ 
are shown as obtained using the Durham jet algorithm.
The numbers of selected udsc, b-tag and gluon jets are shown in 
Table~\ref{LEP12stat}. The larger number of b-tag jets compared to gluon jets 
is due to the inclusive single jet tag method which allows up to two b-tag 
jets per event. 

\subsubsection{Energy-ordering method}\label{EOM}
\begin{figure}[tb] \centering
\vspace*{1cm}
\epsfig{file=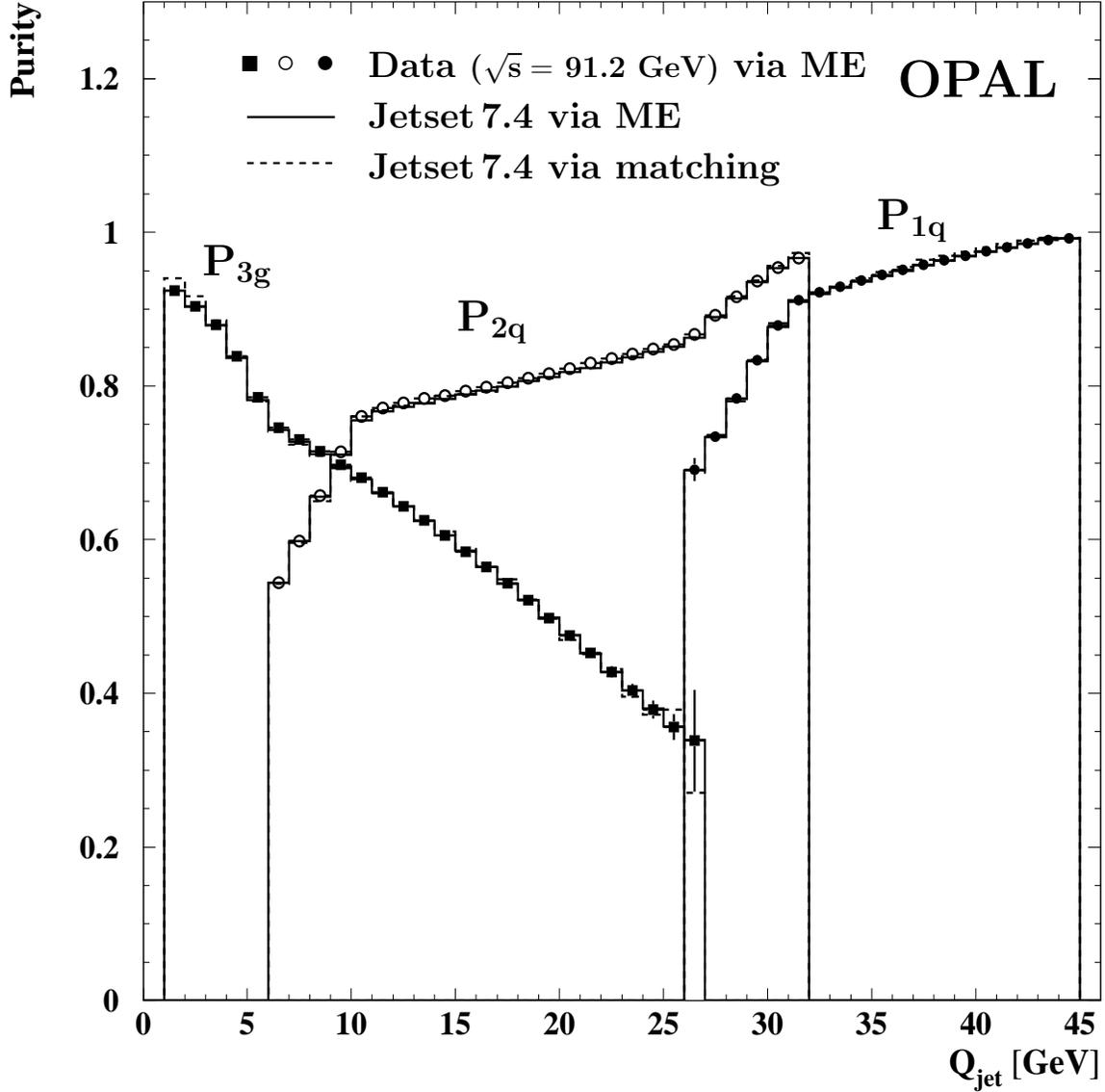,bbllx=82pt,bblly=219pt,bburx=510pt,%
bbury=641pt}
\vspace*{0.5cm}
\caption{The quark purities of jet 1 and 2 samples ($\mr P_{1q}$ and 
$\mr P_{2q}$) and the gluon purity of jet 3 sample ($\mr P_{3g}$) as a 
function of the scale \Qj\ using the Durham jet algorithm. Either the matrix 
element (ME) information (data as symbols and JETSET 7.4 as the solid 
histograms) or the matching method (JETSET 7.4 as dashed histograms) is used. 
Only statistical uncertainties are shown.}
\label{pur-eo1}
\end{figure}

This method is based on the QCD prediction that in a three-jet event the 
lowest energy jet has the highest probability to arise from a gluon. 
In this method only jets 2 and 3 are used, which form the quark and gluon jet 
samples, respectively. There are two ways of estimating the 
purities: either via the matching which employs the inter-jet angles as 
described in the b-tag method, or using matrix elements. It has been shown 
\cite{pur-me} that, for leading order QCD matrix elements, the probability for
a given jet $i$ among the jets \{$i,j,k$\} to be a gluon jet can be expressed 
as a function of the jet energies:
\begin{equation}
P_{i{\mr g}}\propto \frac{x_j^2+x_k^2}{(1-x_j)(1-x_k)}
\label{pur-me}
\end{equation}  
where $x_i=2E_{{\mr jet},i}/\sqrt s$. The corresponding probability for the 
jet to be a quark jet is
\begin{equation}
P_{i\mr q}=1-P_{i\mr g},
\end{equation}
normalised such that $P_{1\mr q}+P_{2\mr q}+P_{3\mr q}=2$. Thus, in 
this way, the purities can be obtained based on the kinematics of the data, 
without recourse to MC information. The scale dependence of the quark purities 
of jets 1 and 2, and the gluon purity of the jet 3, are shown in 
Fig.~\ref{pur-eo1}. Good agreement is obtained 
between the data and MC for the matrix element method. The MC results based on 
matching are seen to agree well with the results based on the matrix elements.
For consistency reasons, the purities based on the matching are used to obtain
the final results. \\

An unfolding to the level of pure quark and gluon jets is 
carried out by solving the following equation:
\begin{equation}
\stackrel{\mr \ \ \ \ \ uncor}{
\left( \begin{array}{l}D_2 \\
                       D_3 \end{array} \right)\!(x_{\mr E},Q)} \ \ = \ \ 
\left( \begin{array}{lll}P_{\mr 2q}~\:P_{\mr 2g} \\
                         P_{\mr 3q}~\:P_{\mr 3g} \end{array} \right)\!(Q)
 \! \stackrel{\mr \ \ \ \ pure}{
\left( \begin{array}{l}D_{\mr q} \\
                       D_{\mr g} \end{array} \right)\!(x_{\mr E},Q)}\\
\label{qgeo}
\end{equation}
where $D_{2(3)}^{\mr uncor}$ is the detector level distribution of a variable 
$D$ in the sample of jets 2 (3) and $D_{\mr q(g)}^{\mr pure}$ corresponds to
pure quark (gluon) jets. The energy-ordering 
method can only be applied in the \Qj\ region where the samples of 
jets 2 and 3 overlap ($6<Q_{\mr jet}<$ 27~GeV for the LEP1 and 
$10<Q_{\mr jet}<$ 60~GeV for the LEP2 sample). 

\subsubsection{Hemisphere method}\label{hemisphere}
In the inclusive hadronic event sample we again use b-tagging to obtain 
samples enriched in b and udsc jets. In the LEP1 sample, a b-tag event is 
defined by requiring two secondary vertices with VNN$>$0.8, while in the LEP2 
sample---due to limited statistics---only one secondary vertex with VNN$>$0.8 
is required. All remaining events form the udsc event sample. Events with no
requirement on the presence of a secondary vertex form the inclusive hadronic
event sample. Each event contains two unbiased jets (hemispheres) of the same 
energy, $\sqrt s/2$. The jets are unfolded to the level of pure udsc
and b jets using an analogous procedure to that described in Section \ref{EOM}
for the energy-ordering method. In Eq.~(\ref{qgeo}), we replace the indices 2 
and q by the index l, and the indices 3 and g by the index b. The purity
$P_{\mr bl}$ ($P_{\mr ll}$) then denotes the probability that a jet from the 
b-tag (udsc) jet sample comes from an u,d,s or c-quark. 
In the LEP1 MC sample, $P_{\mr ll}=$79\% and 
$P_{\mr bb}=$99.7\% which means that we work with a very pure b-tag jet sample.
The corresponding purities for the LEP2 MC sample are 
89\% and 75\%. The numbers of unbiased udsc and b-tag jets passing the 
selection cuts together with background estimates (for the LEP2 data) are 
summarized in Table~\ref{LEP12stat}. The higher efficiency of selecting 
non-ISR events and the lower background compared to those for the biased jets 
is due to the tightened cut on the c.m.s.\ energy described in 
Section~\ref{evesel}. 

\subsection{Construction of flavour inclusive \ff\ from \\
biased jets}\label{ficor}
To construct the flavour inclusive \ff\ from the LEP1 biased jets, the samples
of udsc, b-tag and gluon jets from the b-tag method are used. The quark jet 
sample is formed by a sum of the udsc and b-tag jet samples. The unfolding to 
the level of pure quark and gluon jets can then proceed by use of 
Eq.~(\ref{qgeo}) where the sample of jets 2 is replaced by the quark jet 
sample and the sample of jets 3 by the gluon jet sample. 
To take into account the 
$\sqrt s$ dependence of the flavour composition of the primary \qq\ pair, the 
sample of pure quark jets is constructed as a sum of samples of pure udsc and
b jets, weighted by factors of $r_{\rm udsc}(\langle Q_{\mr jet}\rangle)$ 
and $r_{\mr b}(\langle Q_{\mr jet}\rangle)$, respectively. The 
$r_{\mr b}(\langle Q_{\mr jet}\rangle)$ factor is calculated using the hadron 
level MC, as the ratio of the b$\bar{\mr b}$ production rate for a given
$Q_{\mr jet}$ bin with a mean value of $\langle Q_{\mr jet}\rangle$ in 
three-jet events generated at $\sqrt s =$ 91.2~GeV and the b$\bar{\mr b}$ 
production rate in inclusive hadronic events generated at 
$\sqrt s = \langle Q_{\mr jet}\rangle$. 
The factor $r_{\rm udsc}(\langle Q_{\mr jet}\rangle)$ is determined in an 
analogous fashion. The corrections based on $r_{\mr b}$ and $r_{\rm udsc}$ are
smaller than 15\% and bring the biased jet data closer to the published 
unbiased jet data.  

\subsection{Correction procedure}\label{correction}
The remaining 4-fermion background in the LEP2 data is estimated for each
observable by MC simulation and subtracted on a bin-by-bin basis from the
data distributions, as already mentioned in Section \ref{evesel}. Then the data
and MC distributions at the detector level 
are corrected to the level of pure quarks and gluons by solving either 
Eq.~(\ref{purmat}) or Eq.~(\ref{qgeo}).
As a  last step, we correct the data for the effects of limited detector 
acceptance and resolution as well as for the presence of remaining radiative 
events. The data are multiplied, bin-by-bin, by correction factors calculated
as ratios of distributions at the hadron level to those at the detector level.
For the hadron level biased jets, the same jet selection 
criteria as described in Section \ref{jetsel} are applied except that the
jets are not required to satisfy $|\cos\theta_{\mr jet}|\le 0.90$. 
The quark and gluon jets at the hadron level are identified with MC 
information using the matching technique described in Section 
\ref{btag}. The correction factors from JETSET/PYTHIA used to correct the data
do not exceed 20\%. The correction factors from HERWIG used to estimate the
model dependence of the results are similar. A bin-by-bin correction procedure
is suitable for the measured distributions as the detector and ISR effects do
not cause significant migration (and therefore correlation) between bins.
Typical bin purities for the \Qj\ binning chosen were found to be 75\%, 
the lowest value was 65\%.  

\subsection{Systematic uncertainties}  
The systematic uncertainties of the measurements are assessed by repeating the
analysis with the following variations to the standard analysis.\\

\begin{enumerate}

\item The systematics on the modelling of the Z and Z$^*/\gamma^*$ events used 
to correct the data for ISR, detector effects and quark and gluon jet 
misidentification is estimated by using HERWIG instead of JETSET/PYTHIA.
In the bulk of the measured data, the maximum differences for all types of 
\ffs\ do not exceed 6\%. In the last \xe\ bin of both types of flavour 
inclusive \ffs\ ($0.8<x_{\mr E}<1.0$), the two models deviate from each other 
by as much as 50--60\%. 

\item To assess any inadequacies in the simulation of the response of the 
detector in the endcap regions, the analysis was restricted to the 
barrel region of the detector, requiring the tracks and electromagnetic
clusters to lie within the range $|\cos\theta_{\mr particle}|<0.70$. 
The maximum differences reach 10\% for biased jets (for large \xe)
and 2\% for unbiased jets.  

\item Potential sensitivity of the results to details of the track selection
is assessed by repeating the analysis with modified track selection criteria:
the maximum allowed distance of the point of closest approach of a track to the
collision point in the $r-\phi$ plane, $d_0$, is changed from 5 to 2~cm, the
maximal distance in the $z$ direction, $z_0$, from 25 to 10~cm and the minimal
number of hits from 40 to 80. The quadratic sum over the deviations from the
standard result, obtained from each of these variations, is included to the 
total systematic uncertainty. In most of the bins, the changes are below 1\%.
Larger changes are observed for high \xe, where they are within 7\% for both,
the biased and unbiased jets. 
 
\item The jet algorithm dependence of the biased jet results is estimated by 
repeating the analysis using Cambridge and cone jet algorithms. The largest 
of the two deviations from the standard result (the cone algorithm in most of 
the bins) is taken as the systematic uncertainty. All differences are within 
10\% for all types of \ffs, except at low \Qj\ and 
\xe\ ($4<Q_{\mr jet}<9$~GeV with $0.02<x_{\mr E}<0.04$) where 
the results of the cone algorithm are about 20\%, 24\%, 31\% and 36\% below 
the results of the Durham algorithm, for the flavour inclusive, udsc, b and 
gluon jet \ff, respectively. The differences between the results for 
individual jet algorithms diminish with increasing jet energy. 

\item The jet selection criteria were varied. The minimum particle multiplicity
per jet is changed from 2 to 4; the minimum corrected jet energy is changed 
from 5 GeV to 3 and 7 GeV; the minimum inter-jet angle is changed from 
30$^{\circ}$ to 25$^{\circ}$ and 35$^{\circ}$ and the minimum sum of inter-jet
angles is changed from 358$^{\circ}$ to 356$^{\circ}$ and 359$^{\circ}$. 
The largest deviation with respect to the standard result is taken as the 
systematic uncertainty. The differences are below 2\% in all cases, except for 
large \xe\ with small \Qj\ where they reach 6\%.  

\item The dependence of the results on the neural network output value is 
estimated by varying the cut on VNN from 0.50 to 0.95. The maximum of the 
deviations with respect to the standard result is taken as the systematic 
uncertainty. Typical deviations are 2\% for unbiased jets and the LEP1 biased 
jets, while they are 5\% for the LEP2 biased jets. 
The largest deviation is 11\% for the unbiased jets and 20\% for 
the biased jets (both observed for large \xe). 

\item The b-tagging efficiency is determined using MC events. The systematic
uncertainty in this efficiency was estimated to be about 5\% for VNN$>$0.50
in LEP2 data \cite{Higgs}. The effect of this uncertainty is assessed by
changing the VNN thresholds in the MC samples such that the b-tagging 
efficiency increases or decreases by 10\%, while leaving the thresholds in the
data unchanged. The largest deviation with respect to the standard result is 
taken as the systematic uncertainty. In most of the bins, the differences are 
below 1\%. In the high \xe\ region, they reach 4\% for unbiased jets and are 
typically within 8\% for biased jets. 

\item The uncertainty in the estimates of purities for the b-tag method is 
accounted for by using the non-flavour assignment instead of the flavour 
assignment of the outgoing primary \qq\ pair and gluon to three parton jets. 
Non-negligible differences in the purities are seen only in those 
\Qj\ regions where the purities are small. This results in negligible
effects on the final results: they are below 1\% everywhere. In case of the 
energy-ordering method, the procedure based on the matrix elements is used
instead of the matching. The differences for the gluon jet \ffs\ are below 1\%
everywhere. 

\item Uncertainties arising from the selection of non-radiative LEP2 events 
are estimated by using the ``energy balance'' procedure instead of the
``invariant mass'' procedure. The differences are below 5\% for both
the biased and unbiased jets. 

\item Systematic uncertainties associated with the subtraction of the 4-fermion
background events in the LEP2 samples are estimated by varying the cut on 
$W_{\mr QCD}$ from -0.5 to 0.0 and -0.8. The maximum of the deviations with 
respect to the standard result is taken as the systematic uncertainty. 
The differences are below 4\% for both the biased and unbiased jets. 
In addition, we varied the predicted background to be subtracted by $\pm$5\%, 
slightly more than its measured uncertainty at $\sqrt s = 189$ GeV of 4\% 
\cite{bgr_189}. The differences are below 1\% everywhere.

\end{enumerate}

The results for the udsc jets are found to be less sensitive to the 
above variations than the results for b and gluon jets. The largest 
changes in the numbers of selected b and gluon jets relative to those 
shown in Table~\ref{LEP12stat} are given by variation 6. For the LEP1 sample,
the number of b-tag (gluon) jets grows by 55\% (44\%) for VNN=0.5 and 
drops by 40\% (37\%) for VNN=0.95. Variation 6 also gives rise to the 
most significant change in the purities of the b-tag and gluon jet samples. 
The b (gluon) purity decreases by 17\% (5\%) for VNN=0.5, while it increases
by 7\% (2\%) for VNN=0.95 (the b-purity shown in Fig.~\ref{pureff12}a). 
Other variations change the purities very little. \\

The differences between the standard results and those found using each of 
the above conditions are used to define symmetric systematic uncertainties.
To reduce the influence of statistical fluctuations, the systematic 
uncertainties from all sources are determined for a few larger \Qj\ bins, 
each of them exactly covering two or three original bins. The systematic 
deviation found for this larger bin is then assigned to all original bins 
contained in it. The total systematic uncertainty is defined as the quadratic 
sum of these deviations.

\section{Monte Carlo comparison of biased and unbiased jets}\label{bias}
As discussed above, jets found using a jet algorithm are biased and in this
sense are less suitable for comparison with theory than unbiased jets.
To assess the difference between biased and unbiased jets, we perform a
comparison of their properties using hadron level MC event samples. 
For this purpose, we choose HERWIG because it contains an event generator for 
gg events from a colour singlet point source and because it describes well 
gluon jet properties \cite{gincl3}.\\

\begin{figure}[tb] \centering
\epsfig{file=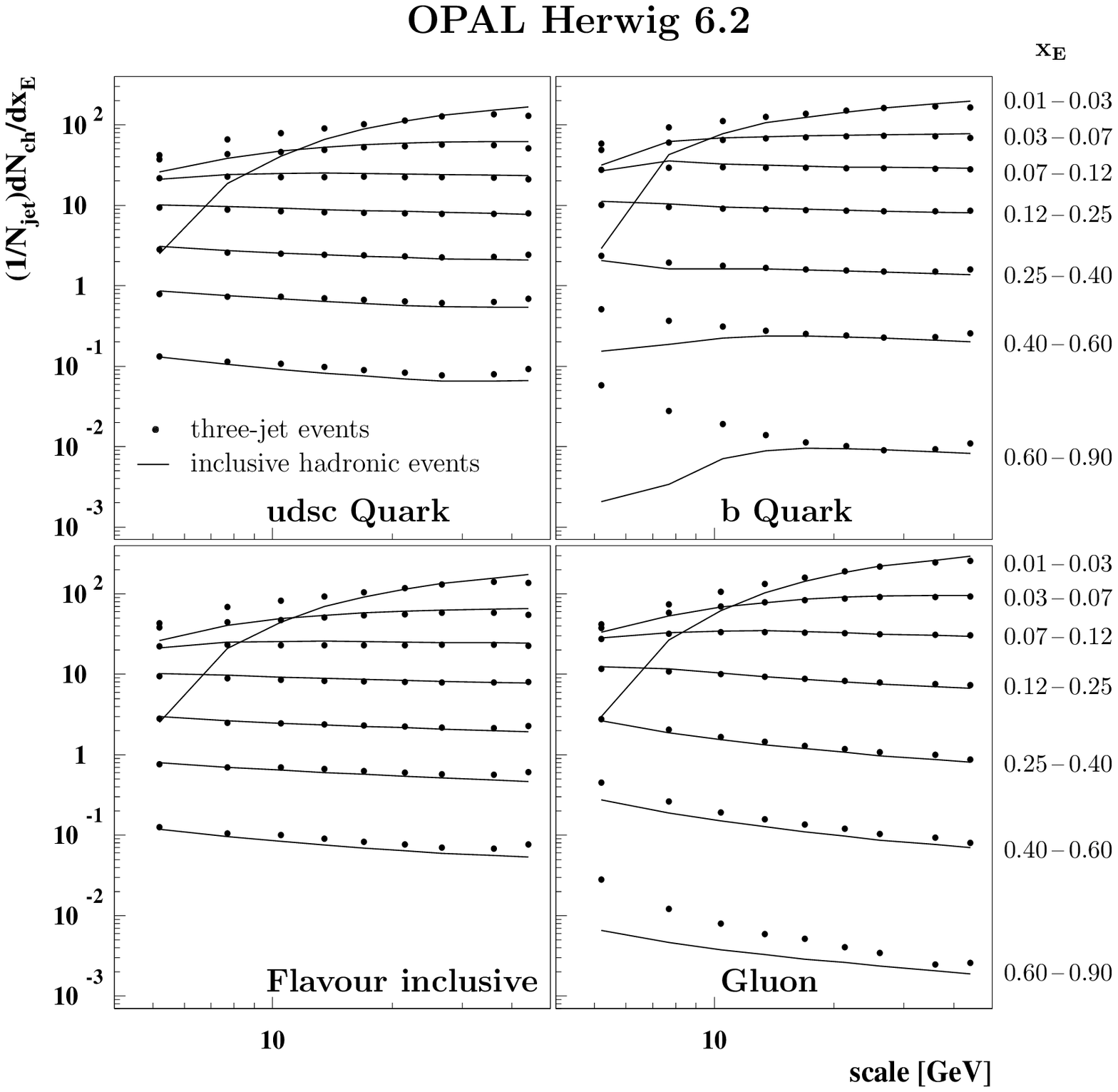,bbllx=58pt,bblly=210pt,bburx=543pt,%
bbury=686pt,width=\textwidth,height=\textwidth}
\caption{The \ffs\ in bins of \xe\ and scale as obtained from hadron
level events generated with HERWIG 6.2. The scale stands for \Qj\
in the case of biased jets (generated at $\sqrt s =$ 91.2~GeV) and for 
$\sqrt s/2$ in the case of unbiased jets. 
The Durham algorithm is used to find jets. The statistical uncertainties 
are smaller than the symbols.}
\label{bias-unbias}
\end{figure}

The conclusions from the comparison of the biased and unbiased jet \ffs\ are 
basically independent of the Monte Carlo model and jet algorithm used in the 
analysis, therefore, as an example, we show in Fig.~\ref{bias-unbias} the 
comparison for HERWIG 6.2 and the Durham jet algorithm. The results correspond
to the hadron level described in Section \ref{correction}. The three-jet 
events (i.e. containing biased jets) are generated at $\sqrt s =$ 91.2~GeV. 
The inclusive hadronic events (no jet finder used, so containing unbiased jets)
are generated separately at values of $\sqrt s$ corresponding to twice the 
central values of \Qj\ in the individual \Qj\ intervals used in the analysis 
of three-jet events. 
Differences between biased and unbiased jet properties are expected due to
different scales used (\Qj\ vs. $\sqrt s/2$) and different number of
jets per event (two hemispheres vs. three jets found by a jet algorithm and
spatially restricted by the minimum inter-jet angle of 30$^{\circ}$).
We point out four regions of phase space where the differences between the 
biased and unbiased jet \ffs\ are larger than 15\%:
\begin{description}
\item a) {\sl Small scales with small \xe\ for all \ffs}: 
This difference, which decreases with increasing scale and \xe, may in part be 
explained by hadron mass effect. At small c.m.s.\ energies, hadron masses are 
not negligible with respect to jet energies, causing a suppression of the 
\ffs\ at very low \xe. This effect is not present in theory (hadrons are taken
to be massless) and is less strong in three-jet events (the mean value of 
$E_{\mr jet}$ in the first \Qj\ bin 
is about 13~GeV) so one can expect the three-jet data to be better 
described by theory than the unbiased jet data in these bins. 

\item b) {\sl Small scales with large \xe\ for b jet \ffs}: 
Since this difference increases with increasing \xe\ and decreasing 
scale, it might be explained by the b-quark mass effect, i.e. by the ratio 
$m_{\mr b}/E_{\mr jet}$. At small c.m.s.\ energies, just above the 
${\mr b\bar{b}}$ production threshold ($\sqrt s = 2m_{\mr b} \approx 10$~GeV),
the above ratio is close to 100\% and almost all particles picked up in the 
hemispheres come from decays of B hadrons, resulting in a very small 
probability to produce a particle with \xe\ close to unity. 
As the scale increases, there is more and more phase space for gluon showers, 
leading to more hadrons from the fragmentation process. However, the number of
radiated gluons is limited by the so-called ``dead cone 
effect'' \cite{deadcone}, i.e. by a suppression of the gluon emission within an
angle of order $m_{\mr b}/E_{\mr jet}$. In three-jet events, 
the ratio $m_{\mr b}/E_{\mr jet}$ starts at a much smaller value than
in hemisphere events (since the mean jet energy in the first \Qj\ bin is about
13~GeV) leading to much more phase space for gluon showering compared
to hemisphere jets with the same value of scale. 
In QCD calculations based on unbiased jets, this ratio can be identified 
with mass terms of the type $m_{\mr q}/Q$ where $Q$ is 
some hard scale. In current NLO calculations, these mass terms are not 
considered. As will be seen later, the three-jet data and theory behave 
similarly in the region of small scales. This similarity suggests that 
missing mass terms in theory may behave like $m_{\mr b}/E_{\mr jet}$. 

\item c) {\sl Large \xe\ for gluon jet \ffs}: The sizable discrepancy 
observed for $x_{\mr E}>0.6$ clearly suggests a bias in the gluon jet 
results. It appears to be more appropriate \cite{unbiasgluon} to 
consider for example both the energy scale and the exact virtuality scale and 
to boost to a frame in which the two scales are equal. MC studies recently 
presented by OPAL in \cite{indgl2} demonstrate that such boosted gluon jets 
are less biased than those from our study, in particular in the regions of 
very small and large \xe.

\item d) {\sl The last scale bin for all quark jet \ffs}: The observed 
difference is larger than 15\% in the \xe\ ranges of 0.01--0.07 and 0.40--0.90.
Although biased jets in the interval $44<Q_{\mr jet}<46$ GeV
should in principle resemble hemispheres of the same energy (due to large 
angles $\vartheta$ reaching up to 165$^{\circ}$), we found that the 
soft particle multiplicity differs between the two cases. Therefore this 
difference is considered to represent a true bias of biased jets. 

\end{description}

The comparisons made in this MC study suggest that biased jets are less 
sensitive to hadron and b-quark mass effects than unbiased jets. This implies
that biased jets tend to be more appropriate for comparisons with
theory than unbiased jets in the regions of low scale with low \xe,
and in case of b jets, also at low scale with high \xe.

\section{NLO predictions}\label{NLO}
The results are compared to theoretical predictions by three groups, namely
Kniehl, Kramer and P\"{o}tter (KKP) \cite{KKP}, Kretzer (Kr) \cite{Kretzer} 
and Bourhis, Fontannaz, Guillet and Werlen (BFGW) \cite{BFGW}. The three 
groups provide numerical values of the quantity defined in Eq.~(\ref{ffdef}), 
up to the next-to-leading order in $\alpha_S$. This means that in the 
extraction of these predictions from measured charged particle momentum 
distributions, the hard scattering cross section for the production of a 
parton in
\ee\ annihilation is evaluated to an accuracy of the order $\alpha_S$, while 
the splitting functions describing the scale dependence are evaluated to an 
accuracy of the order $\alpha_S^2$. We stress that these NLO predictions
correspond to an unbiased jet definition. The scale evolution via DGLAP 
evolution equations is performed starting from fragmentation functions at a 
fixed input scale, extracted from existing measurements. In each of these 
calculations, the renormalization and fragmentation scales are set equal to 
the hard scale $Q$. The calculations, nevertheless, differ in a number of 
important aspects, such as the choice of data sets, the definition of the scale
$Q$, the fit ranges, the prescription for the number of active flavours in 
the evolution of \ffs\ and partonic cross sections, and the treatment of heavy 
flavours and gluons. \\

More specifically, in \cite{KKP} the evolution of the b jet \ff\ starts at 
scale $Q=2m_{\mr b}$ where $m_{\mr b}$ is the b-quark mass put equal to 
4.5~GeV. The number of active flavours, $N_{\mr active}^f$, is driven by twice
the quark mass, $2m_{\mr q}$ ($N_{\mr active}^f=4$ for 
$2m_{\mr c}<Q<2m_{\mr b}$ and similarly for other flavours). The QCD scale
parameter for five flavours and the $\overline{\mr MS}$ renormalization 
scheme, 
$\Lambda_{\overline{\mr \scriptscriptstyle MS}}^{\scriptscriptstyle (5)}$, is 
set equal to 0.213~GeV. 
In \cite{Kretzer} the start of the b jet \ff\ evolution is at the scale
$Q=m_{\mr b}$, $N_{\mr active}^f$ is driven by $2m_{\mr q}$ and 
$\Lambda_{\overline{\mr \scriptscriptstyle MS}}^{\scriptscriptstyle (5)} 
=$ 0.168~GeV.
In \cite{BFGW} the \ffs\ are evolved using an ``optimal'' scale, $Q_{\mr opt}$,
given by the relation $Q^2 \frac{\delta D_a^h}{\delta Q^2}\left | _{Q=Q_{\mr 
opt}} \right. =0$. 
The evolution of the b jet \ff\ starts at scale $Q=m_{\mr b}$, and 
$\Lambda_{\overline{\mr \scriptscriptstyle MS}}^{\scriptscriptstyle (4)} 
=$0.300~GeV. \\

The predictions for quark jet \ffs\ by KKP, Kr and BFGW were made using data
from \cite{loweren,B-decay,flavourff,TPC2} or similar results. 
Concerning the predictions for gluon jet \ffs, it is important to 
note that in \cite{KKP} a fit was made to the unbiased \cite{gincl3} and 
biased \cite{gl26.2} jet data, in \cite{Kretzer} the 
predictions were obtained from the evolution and the NLO correction to the 
\ee\ cross section and in \cite{BFGW} a fit was made to  large 
$p_T$ charged particle data \cite{largept}. Therefore, the experimental input
for gluon jets is very different in the three calculations. 
The fit ranges used by KKP, Kr and
BFGW were $0.1<x_{\mr E}<1.0$, $0.05<x_{\mr E}<0.8$ and $0.12<x_{\mr E}<0.9$, 
respectively.
We obtained the NLO predictions of Kr and BFGW using the code 
\cite{NLOcode} where they are provided in parameterised forms. The relative 
difference between the parameterisation and the exact evolution for predictions
by Kr are smaller than 3\% and 10\% for $x_{\mr E}<0.75$ and $x_{\mr E}<0.90$, 
respectively. All the NLO curves by KKP shown in this analysis correspond to 
the exact scale evolution.\\

We point out that in the NLO predictions, the NLO (of the order $\alpha_S$) 
corrections to the hard subprocess correspond to inclusive hadron production.
For three-jet events, NLO corrections are not available and are 
expected to depend on the jet algorithm used. Our assumption in this analysis 
is that where the biased jet data are observed to be in a good agreement with 
the unbiased jet data, the unknown NLO corrections are apparently small, and 
the biased jet results can be compared to the existing NLO predictions.
Despite the sizable differences between the biased and unbiased jet MC results
reported in points a) and b) of Section \ref{bias}, the biased jet data at low
scales are still considered to be appropriate for such a comparison for the
reasons mentioned at the end of Section \ref{bias}.  

\section{Results}\label{results}
In the following, the results from this analysis are compared with existing
measurements as well as with various fragmentation models and theoretical NLO
predictions. The \ffs\ are presented either with emphasis on the scale 
dependence or the \xe\ dependence. The scale dependent \ffs\ are plotted in 
several \xe\ intervals as functions of scale. For a given bin of
scale, the data or MC point is placed at the value of the scale at which the 
NLO prediction is equal to its mean value over this bin \cite{stick}. 
An analogous prescription is applied for the \xe\ dependent \ffs. 
Since in the following, the biased and unbiased jet results are often plotted 
on the same figure, we have to accommodate the differences between scale 
definitions and number of jets from which the \ffs\ were extracted. 
Therefore the term scale in the following figures stands for \Qj\ in 
case of biased jets and $\sqrt {s}/2$ in case of unbiased jets. The published
unbiased jet results are scaled by $\frac{1}{2}$ since they refer to the entire
event, thus to two jets. For the NLO predictions, the same prescription as for
the published unbiased jet data is applied. 

\subsection{Scale dependence}\label{scaledep}   
\begin{figure}[tb] \centering
\epsfig{file=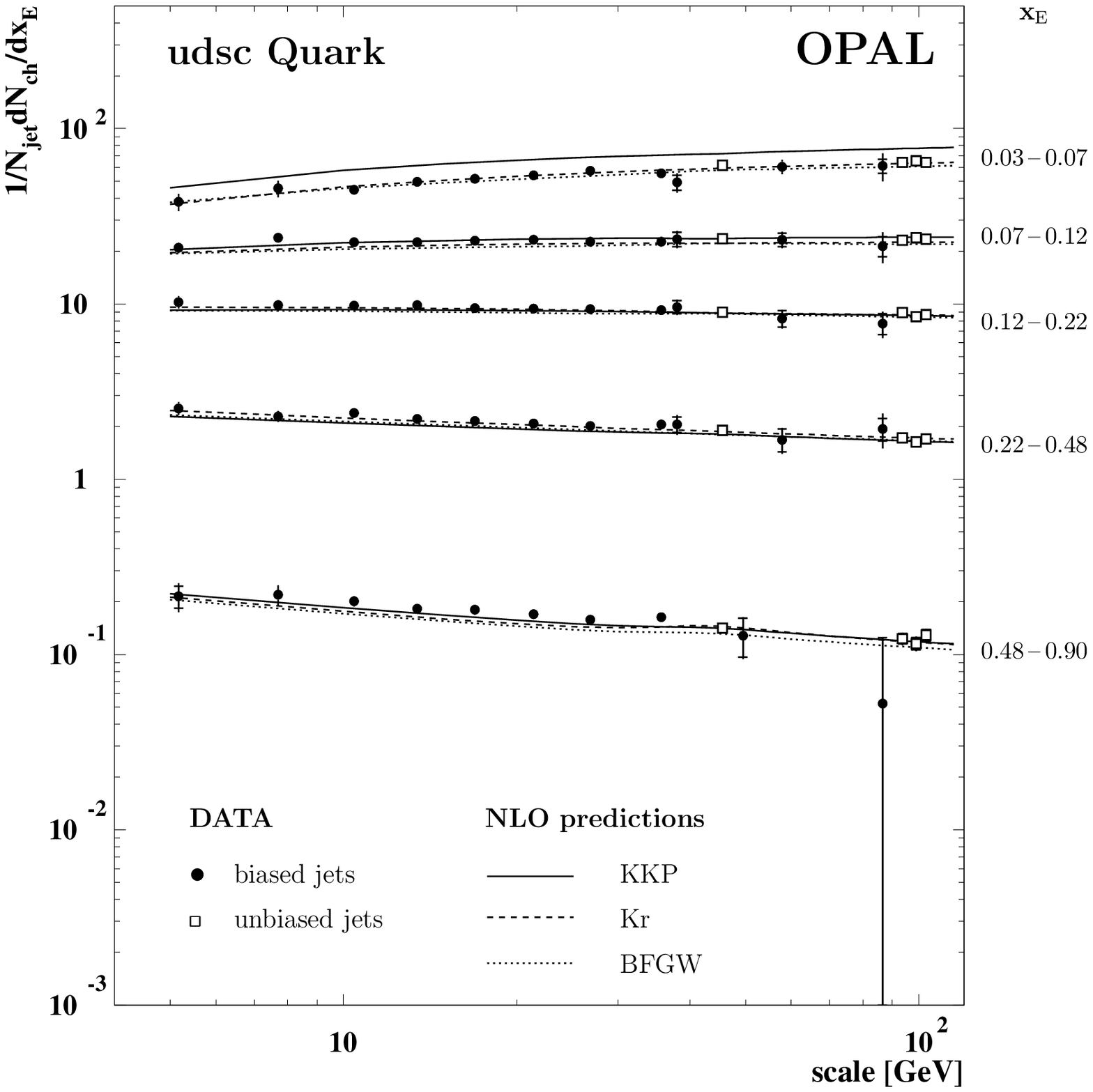,bbllx=58pt,bblly=190pt,bburx=538pt,%
bbury=670pt}
\caption{Scale dependence of the udsc jet \ffs\ in different \xe\ bins.
The scale denotes \Qj\ for the biased jets and $\sqrt s/2$ for the 
unbiased jets. The inner error bars indicate the statistical uncertainties, 
the total error bars show the statistical and systematic uncertainties added 
in quadrature. The values are given in Table~\ref{FFscl}.
The data are compared to the NLO predictions by KKP \cite{KKP}, 
Kr \cite{Kretzer} and BFGW \cite{BFGW}. }
\label{fragl}
\end{figure}

\begin{figure}[tb] \centering
\epsfig{file=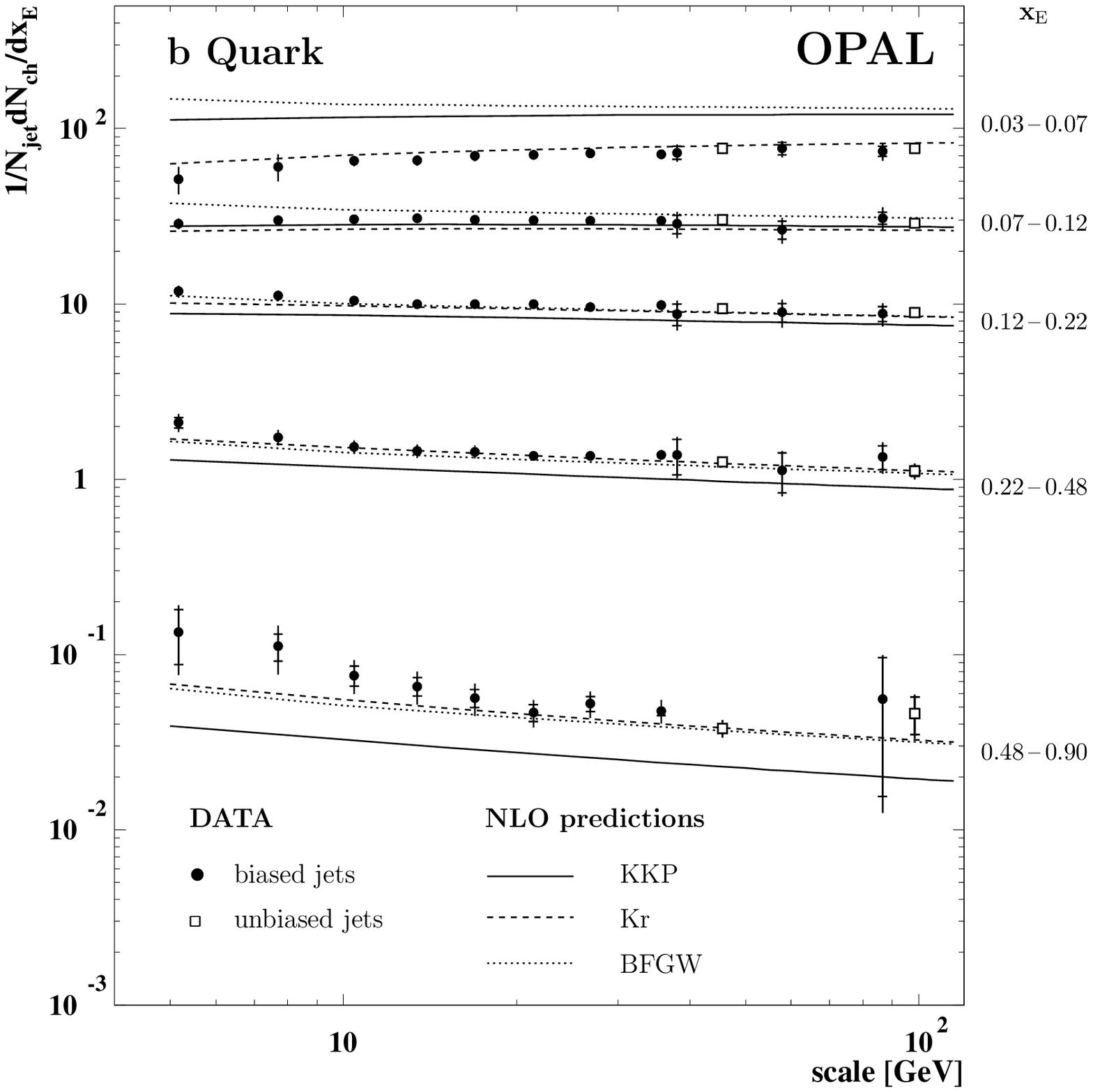,bbllx=58pt,bblly=190pt,bburx=538pt,%
bbury=670pt}
\caption{Scale dependence of the b jet \ffs\ in different \xe\ bins.
The scale denotes \Qj\ for the biased jets and $\sqrt s/2$ for the 
unbiased jets. The inner error bars indicate the statistical uncertainties, 
the total error bars show the statistical and systematic uncertainties added 
in quadrature. The values are given in Table~\ref{FFscb}.
The data are compared to the NLO predictions by KKP \cite{KKP}, 
Kr \cite{Kretzer} and BFGW \cite{BFGW}.}
\label{fragb}
\end{figure}

\begin{figure}[tb] \centering
\epsfig{file=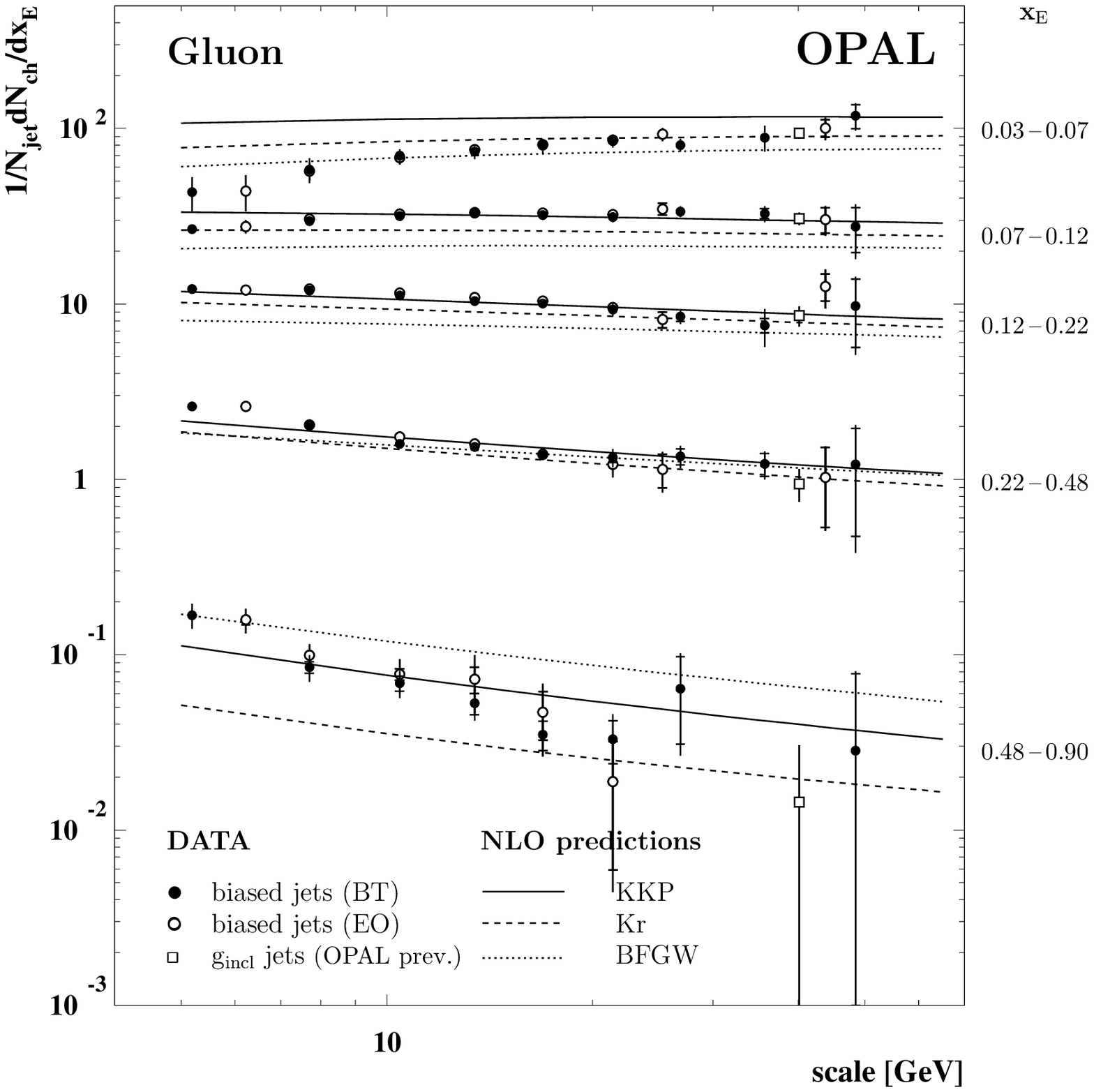,bbllx=58pt,bblly=185pt,%
bburx=538pt,bbury=665pt}
\caption{Scale dependence of the gluon jet \ffs\ in different \xe\ bins.
The scale denotes \Qj\ for the biased jets and $E_{\mr jet}$ for the 
published ${\mr g_{incl}}$ jets (OPAL \cite{gincl3}). The results from the 
biased jets using the b-tag (BT) and the energy-ordering method (EO) are 
shown. The inner error bars indicate the statistical uncertainties, the total
error bars show the statistical and systematic uncertainties added in 
quadrature. The values are given in Table~\ref{FFscg}.
The data are compared to the NLO predictions by KKP \cite{KKP}, 
Kr \cite{Kretzer} and BFGW \cite{BFGW}. }
\label{fragg}
\end{figure}

\begin{figure}[tb] \centering
\epsfig{file=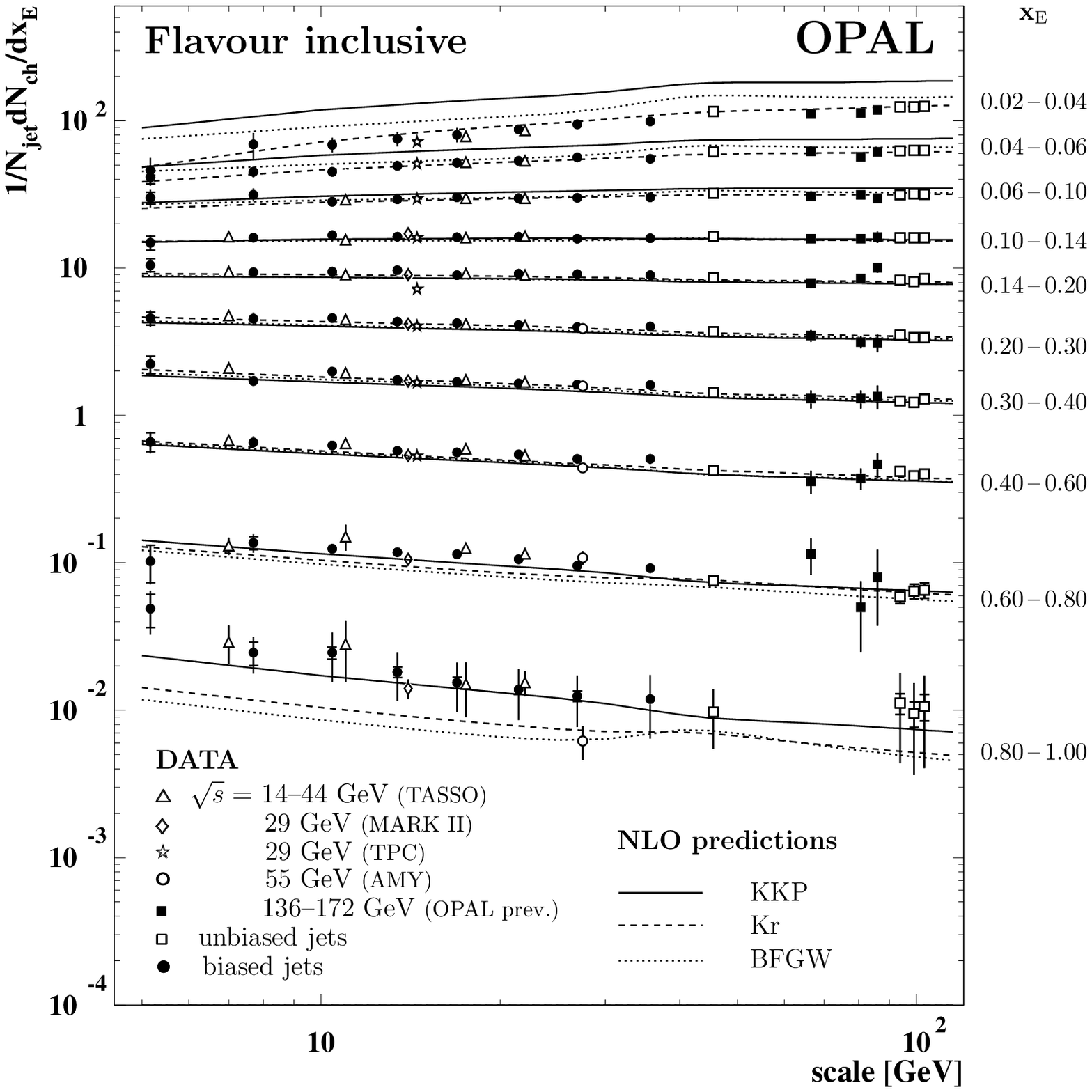,bbllx=57pt,bblly=190pt,bburx=538pt,%
bbury=670pt}
\caption{Scale dependence of the flavour inclusive \ffs\ in 
different \xe\ bins. The scale denotes \Qj\ for the biased 
jets and $\sqrt s/2$ for the unbiased jets. The inner error bars indicate the 
statistical uncertainties, the total error bars show the statistical and 
systematic uncertainties added in quadrature. The values are given in 
Table~\ref{FFincl}.
In addition, the published unbiased jet data by TASSO, TPC, MARK II, AMY 
\cite{loweren} and OPAL \cite{OP133,OP161,OP172} are shown. The data are 
compared to the NLO predictions by KKP \cite{KKP}, Kr \cite{Kretzer} and BFGW 
\cite{BFGW}.}
\label{fraginc}
\end{figure}

In Figs.~\ref{fragl}--\ref{fraginc} and in Tables \ref{FFscl}--\ref{FFincl} 
the results for the udsc, b, gluon and flavour inclusive jet \ffs\ are 
presented. The LEP1 unbiased jet data correspond to $\sqrt s = 91.2$ GeV.
Concerning the LEP2 unbiased jets, the b jet \ffs\ are measured in the entire 
available $\sqrt s$ range of 183--209 GeV. The corresponding data points are 
placed at $\langle \sqrt s \rangle = 197$ GeV, where $\langle \sqrt s \rangle$
is the luminosity weighted value of $\sqrt s$. The udsc and flavour inclusive 
jet \ffs\ are measured in three $\sqrt s$ intervals: 183--189,
192--202 and 204--209~GeV. The corresponding data points are placed at 
$\langle \sqrt s \rangle$ = 187.6, 198.0 and 206.2~GeV, respectively.
The quark biased jet data from LEP1 cover the region $Q_{\mr jet}=$ 4--42~GeV,
while those from LEP2 cover the region $Q_{\mr jet}=$ 30--105~GeV.
The results from the region $0.01<x_{\mr E}<0.03$ are not shown but they are 
discussed in Section \ref{xff}. The results are found to be consistent with 
previous measurements. The \ffs\ from unbiased quark jets agree to within the 
total uncertainties with previous OPAL unbiased jet measurements of flavour 
inclusive and b jet \ffs\ at $\sqrt s=$ 91.2~GeV in \cite{flavourff} and  
flavour inclusive jet \ffs\ at $\sqrt s=$ 192--209~GeV in \cite{Joost} 
(not shown). 
Similarly, the udsc and gluon \ffs\ from biased jets agree
with similar measurements presented by the DELPHI Collaboration \cite{Master} 
for \Qj\ scales between 4 and 30~GeV (not shown). Finally, our gluon 
jet results are seen to be consistent with the results of the ${\mr g_{incl}}$ 
jets \cite{gincl3} at 40.1~GeV, see Fig.~\ref{fragg}. The other results from
our study represent first measurements, specifically the udsc jet results above
45.6~GeV, the gluon jet results above 30~GeV (except for the ${\mr g_{incl}}$ 
jets), and the b jet results at all scales except 45.6~GeV. \\

The data are compared to the theoretical predictions described in Section 
\ref{NLO}. 
For the udsc jet \ff\ (Fig.~\ref{fragl}), all three theoretical predictions 
give a good description in the entire measured phase space, except for 
the lowest \xe\ bin where the KKP calculations overestimate the data, and the
highest \xe\ bin where the data are underestimated by the Kr and BFGW 
calculations. \\

The situation is rather different for the b and gluon jet \ffs\ 
(Figs.~\ref{fragb} and \ref{fragg}) where the description of the data by the 
NLO predictions is worse and where there are significant differences between 
individual NLO results. The latter is, nevertheless, expected due to  
differences in the calculations such as those discussed in Section \ref{NLO}.
In Fig.~\ref{fragb} the KKP prediction is deficient with respect to the data 
for $x_{\mr E}>0.12$. As shown in \cite{B-decay}, with rising particle 
momentum, this region is increasingly populated by the products of B hadron 
decays. It is, however, important to note that these B hadron decay products
are indirectly included in theory predictions since they are present in the
data sets to which the fits were made. \\

For the gluon jet \ffs, the two alternative methods of identifying gluon jets 
described in Section \ref{identif} are examined, see Fig.~\ref{fragg} and 
Table \ref{FFscg}. 
The \Qj\ binning is not the same for the two methods because of their different
regions of applicability. In the LEP1 samples, the interval 
$Q_{\mr jet}=$ 4--42~GeV is used for the b-tag method, while for the 
energy-ordering method, the \Qj\ spectra of jets 2 and 3 overlap in the 
interval $Q_{\mr jet}=$ 6--27~GeV as mentioned in Section \ref{EOM}. In the 
LEP2 samples, the results correspond 
to the interval $Q_{\mr jet}=$ 30--70~GeV for the b-tag method, where only 
jets 2 and 3 are used, and to the interval $Q_{\mr jet}=$ 30--60~GeV for 
the energy-ordering method.
A satisfactory correspondence between the b-tag and energy-ordering methods is
found in the entire scale range accessible. The data tend to show larger 
scaling violations than predicted by any of the calculations. \\

The results for the flavour inclusive jet \ffs\ are presented in 
Fig.~\ref{fraginc} and in Table \ref{FFincl}. 
The results are compared with published unbiased jet data from lower 
energy \ee\ experiments (TASSO, MARK II, TPC and AMY) \cite{loweren} and 
previous OPAL results \cite{OP133,OP161,OP172}. We note that the \ffs\ 
measured by TASSO, MARK II and AMY are defined via $x_{\mr p}=2p/\sqrt s$, 
where $p$ is particle momentum, rather than via \xe\ used in the present
analysis. This difference in definition leads to non-negligible differences 
in the region of $x_{\mr E}<0.1$ and $\sqrt s< 22$~GeV, therefore the 
published data from this region are not shown in Fig.~\ref{fraginc}. The 
results from the current study are seen to be consistent with the previous
results.
The data are also compared to the NLO predictions of KKP, Kr and BFGW. All 
three predictions give a reasonable description of the data in the 
central region of \xe\ ($0.06 \lesssim x_{\mr E} \lesssim 0.60$) and over the 
entire scale range. \\

A good correspondence is found between the results from biased and unbiased 
jets in all four figures. This observation suggests that \Qj\ is an
appropriate choice of scale in three-jet events with a general topology. 
A similar conclusion was previously presented in \cite{Master}. The Monte 
Carlo study described in Section \ref{bias}, however, demonstrates that the 
bias introduced by using jet algorithms in the gluon jet identification is not 
negligible for $x_{\mr E}>0.6$.  
In each of these figures, the scaling violation seen in the data is positive 
for low \xe\ and negative for high \xe. It is more pronounced in the gluon jets
than in the quark jets. \\

\subsection{$\boldmath x_{\boldmath\mr E}$-dependence}\label{xff}
\begin{figure}[tb] \centering
\vspace*{0.5cm}
\epsfig{file=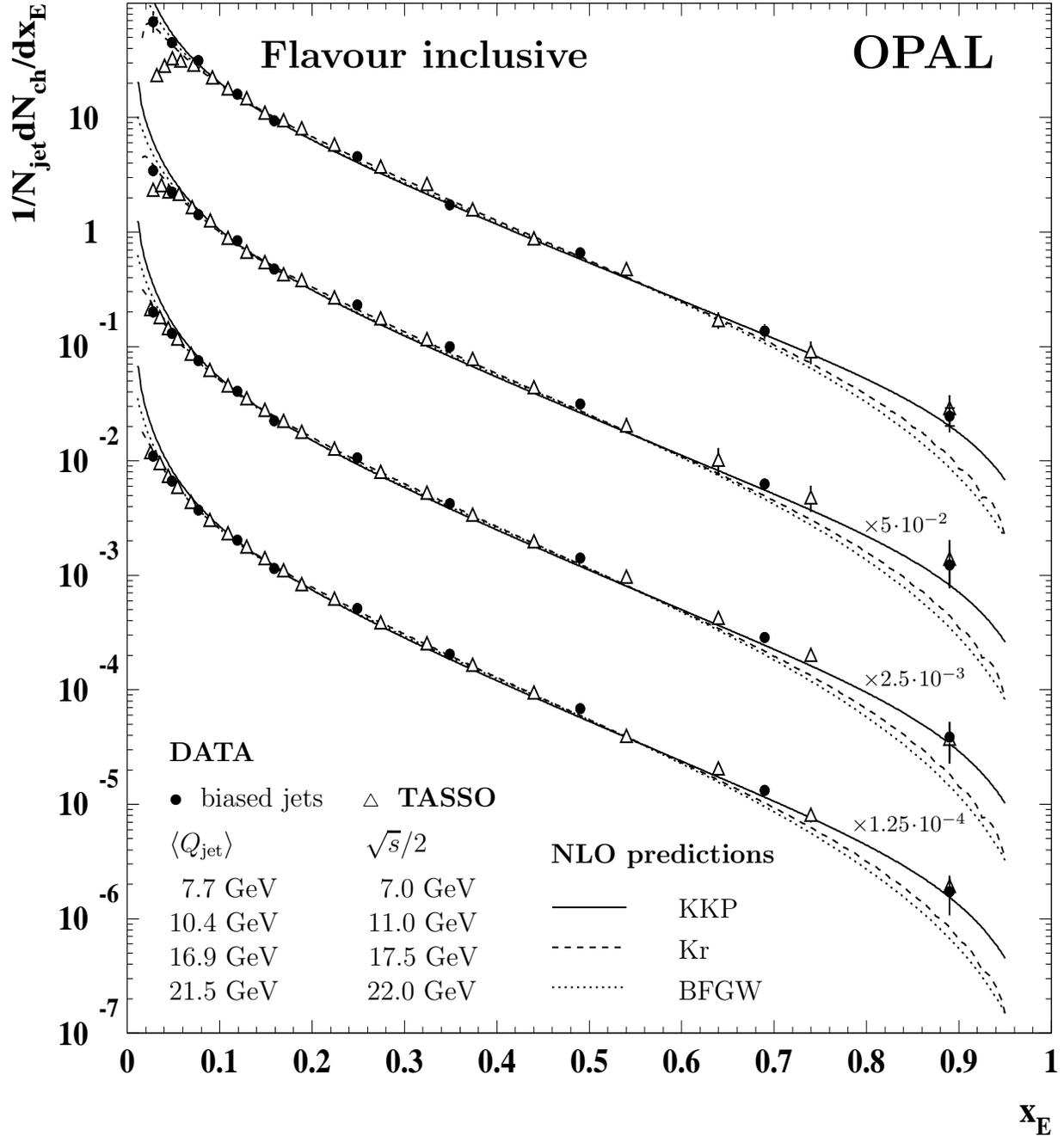,bbllx=68pt,bblly=180pt,bburx=525pt,%
bbury=680pt}
\caption{\xe\ dependence of the flavour inclusive \ffs\ 
at different scales. The inner error bars indicate the statistical 
uncertainties, the total error bars show the statistical and systematic 
uncertainties added in quadrature. The values are given in Table~\ref{FFincl}. 
The results from the biased jets are compared to the results from the unbiased 
jets of TASSO \cite{loweren}. The data are compared to the NLO predictions 
by KKP \cite{KKP}, Kr \cite{Kretzer} and BFGW \cite{BFGW}.
The results are scaled as indicated in figure.}
\label{4sci-nlo}
\end{figure}

In Section~\ref{bias}, we noted the region of small \xe\ with small 
scales where large differences between biased and unbiased jet \ffs\ 
constructed from hadron level MC were observed. In Fig.~\ref{4sci-nlo}, this 
observation is confronted with data. We plot again the unbiased jet data of 
TASSO and the biased jet data from our analysis (Table \ref{FFincl}), the 
latter in those \Qj\ bins which correspond well to the c.m.s.\ energies used in
TASSO measurement. 
We transforme $x_{\mr p} \rightarrow x_{\mr E}$ using the pion mass and 
shifte the TASSO points accordingly. In the first scale bin, the unbiased jet
\ffs\ exhibit turn-over points at very 
low \xe, while the biased jet data grow steeply with decreasing \xe. 
This difference qualitatively confirms the observation we made in point a)
of Section 6 using MC jet samples. Further, as anticipated in 
Section~\ref{bias}, the biased jet data agree better with theory than the
unbiased jet data. \\

\begin{figure}[tb] \centering
\epsfig{file=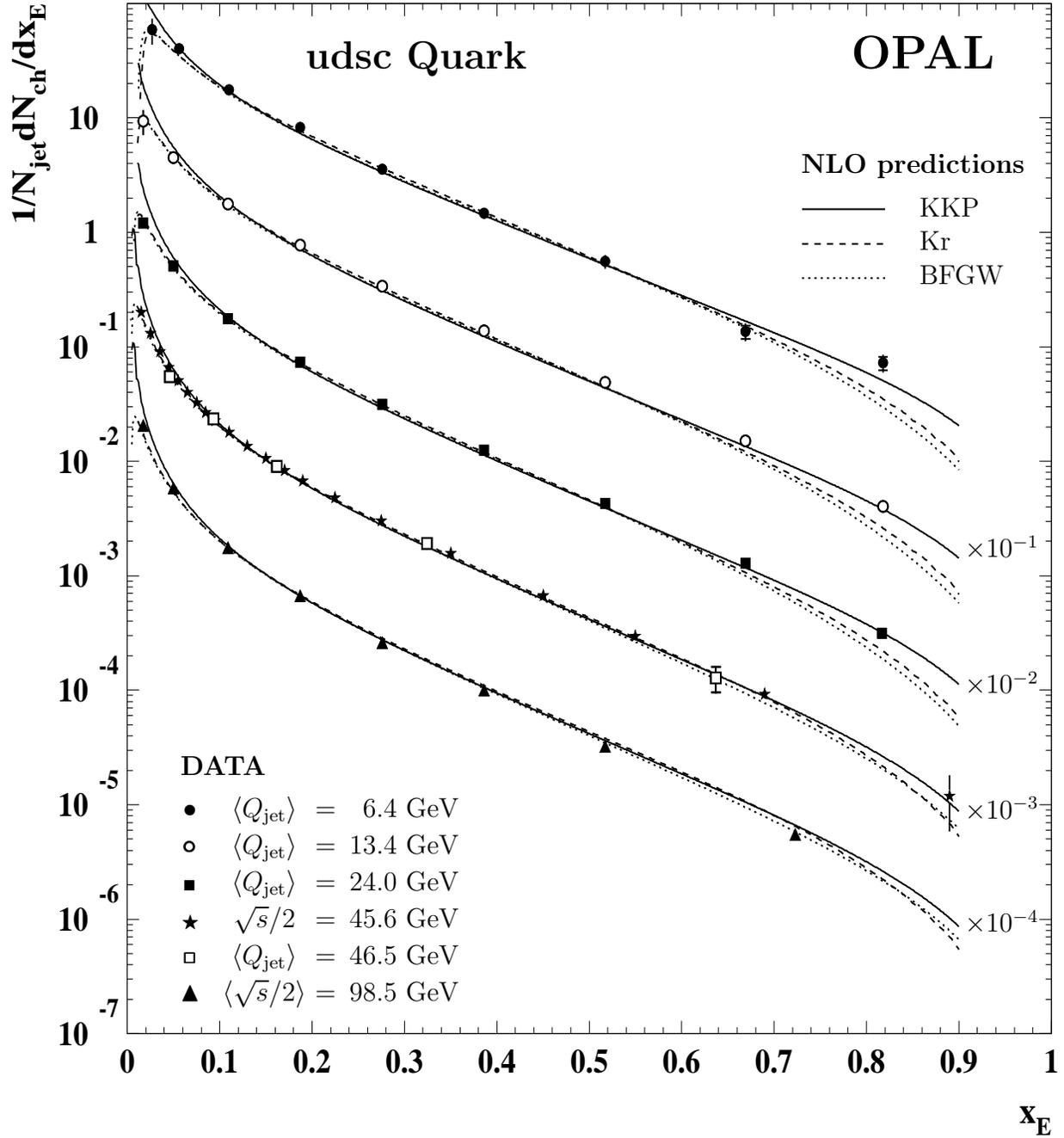,bbllx=68pt,bblly=180pt,bburx=525pt,%
bbury=680pt}
\caption{\xe\ dependence of the udsc jet \ffs\ at different scales. 
The inner error bars indicate the statistical uncertainties, the total error 
bars show the statistical and systematic uncertainties added in quadrature. 
The values are given in Table~\ref{FFx}. The data are compared to the NLO 
predictions by KKP \cite{KKP}, Kr \cite{Kretzer} and BFGW \cite{BFGW}.
The results are scaled as indicated in figure.}
\label{4scl-nlo}
\end{figure}

\begin{figure}[tb] \centering
\epsfig{file=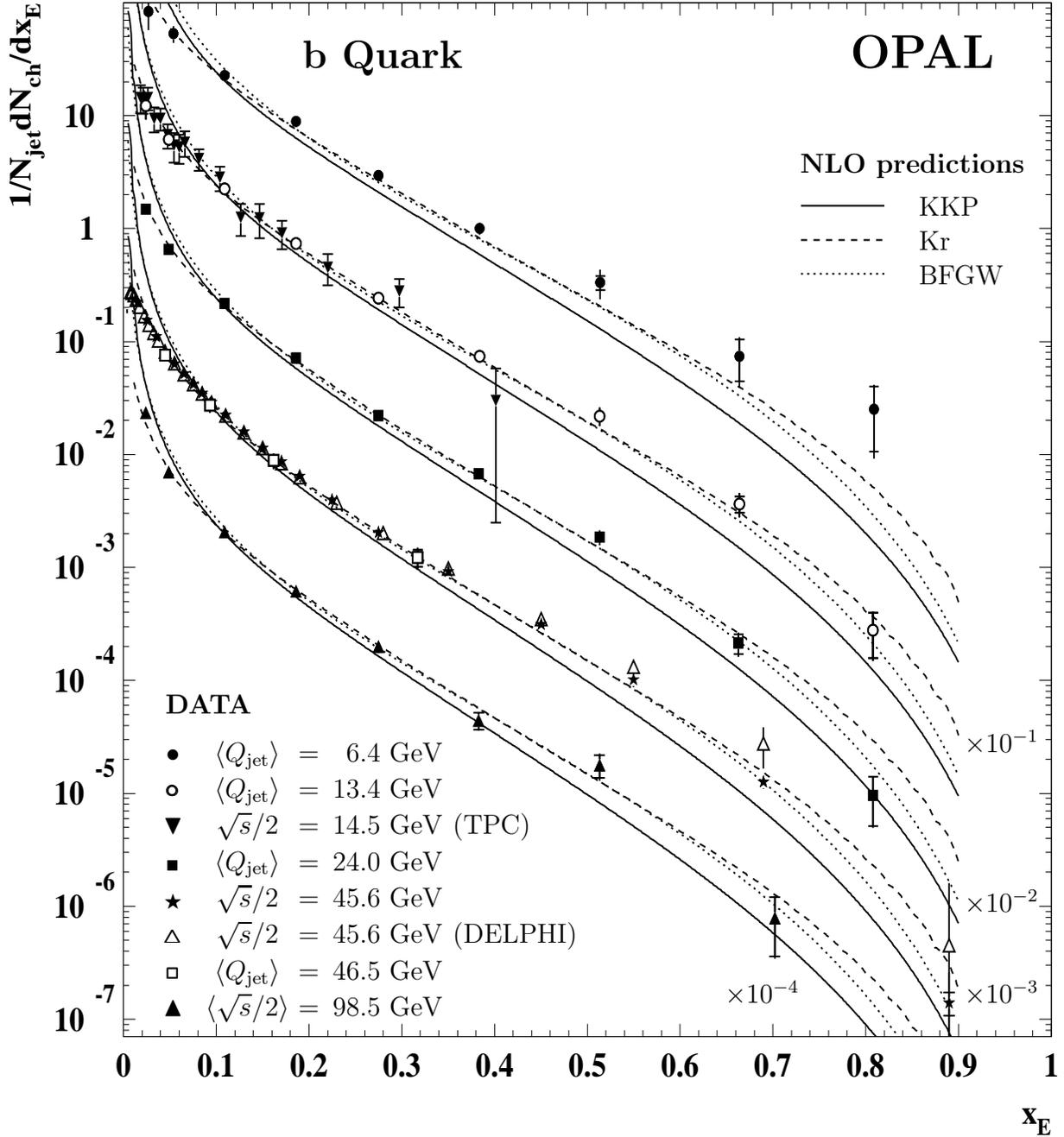,bbllx=68pt,bblly=180pt,bburx=525pt,%
bbury=680pt}
\caption{\xe\ dependence of the b jet \ffs\ at different scales. 
The inner error bars indicate the statistical uncertainties, the total error 
bars show the statistical and systematic uncertainties added in quadrature. 
The values are given in Table~\ref{FFx}. For comparison, the published results
on the unbiased jets of DELPHI \cite{B-decay} and the results based on TPC data
\cite{loweren} are shown. The data are compared to the NLO predictions by KKP 
\cite{KKP}, Kr \cite{Kretzer} and BFGW \cite{BFGW}.
The results are scaled as indicated in figure.}
\label{4scb-nlo}
\end{figure}

\begin{figure}[tb] \centering
\epsfig{file=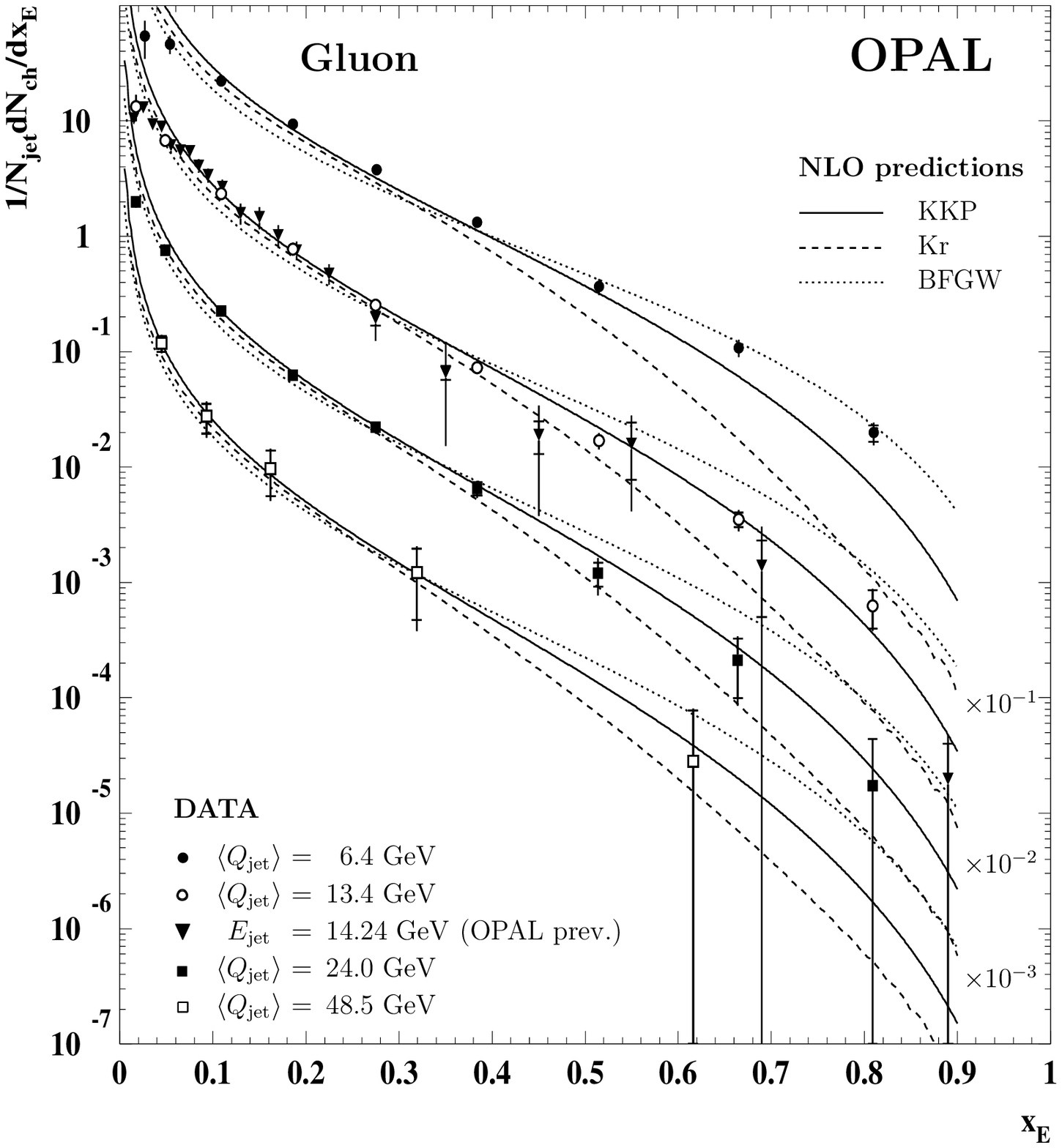,bbllx=68pt,bblly=180pt,bburx=525pt,%
bbury=680pt}
\caption{\xe\ dependence of the gluon jet \ffs\ at different scales. 
The inner error bars indicate the statistical uncertainties, the total error 
bars show the statistical and systematic uncertainties added in quadrature. 
The values are given in Table~\ref{FFx}. Also shown are the recent OPAL 
results on the boosted gluon jets \cite{indgl2}. The data are compared to 
the NLO predictions by KKP \cite{KKP}, Kr \cite{Kretzer} and BFGW 
\cite{BFGW}. The results are scaled as indicated in figure.} 
\label{4scg-nlo}
\end{figure}

\begin{figure}[tb] \centering
\epsfig{file=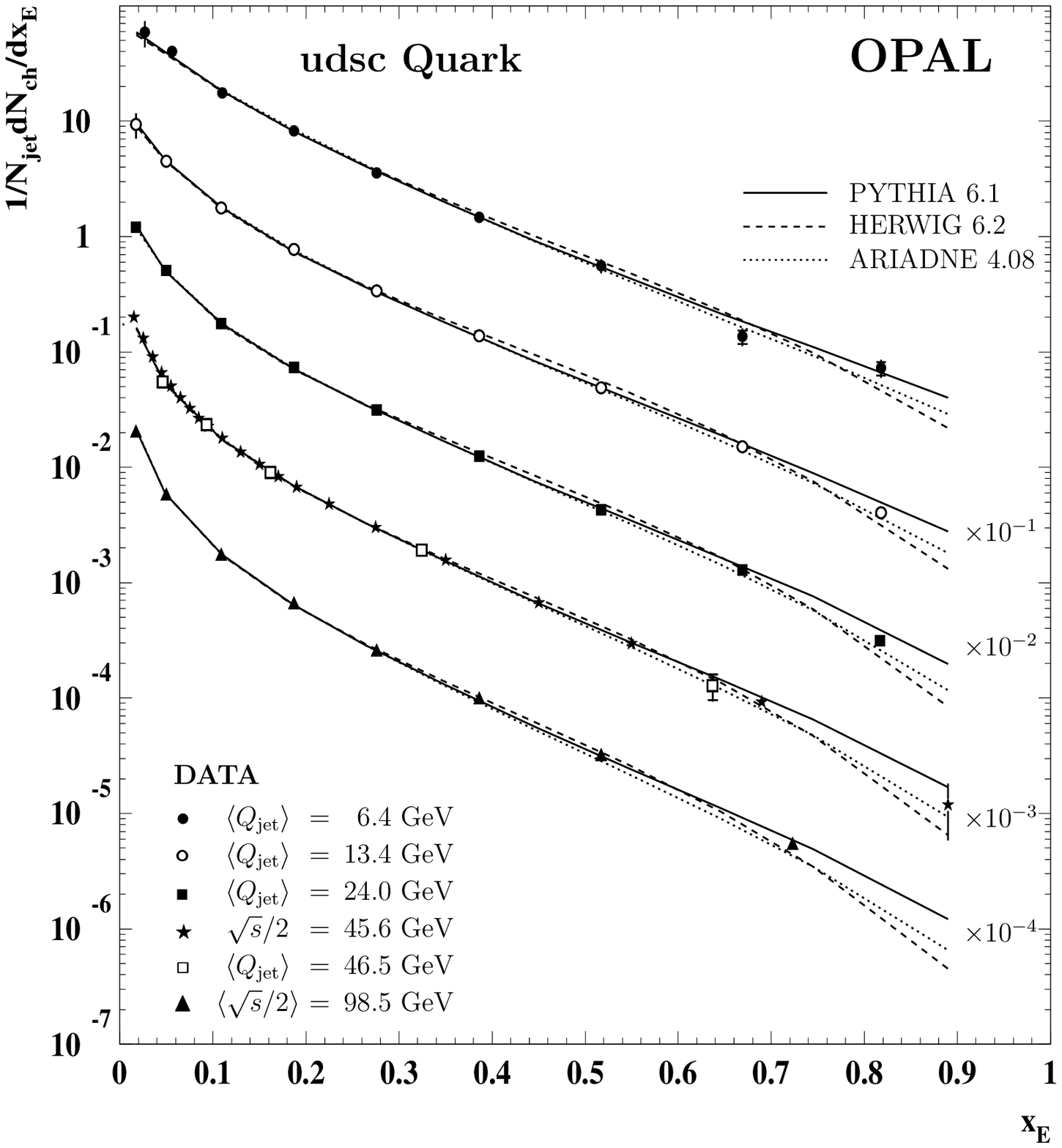,bbllx=68pt,bblly=180pt,bburx=525pt,%
bbury=680pt}
\caption{\xe\ dependence of the udsc jet \ffs\ at different scales 
compared with the PYTHIA 6.125, HERWIG 6.2 and ARIADNE 4.08 Monte Carlo 
predictions. The inner error bars indicate the statistical uncertainties, 
the total error bars show the statistical and systematic uncertainties added 
in quadrature. The values are given in Table~\ref{FFx}.
The results are scaled as indicated in figure.}
\label{4scl-mod}
\end{figure}

\begin{figure}[tb] \centering
\epsfig{file=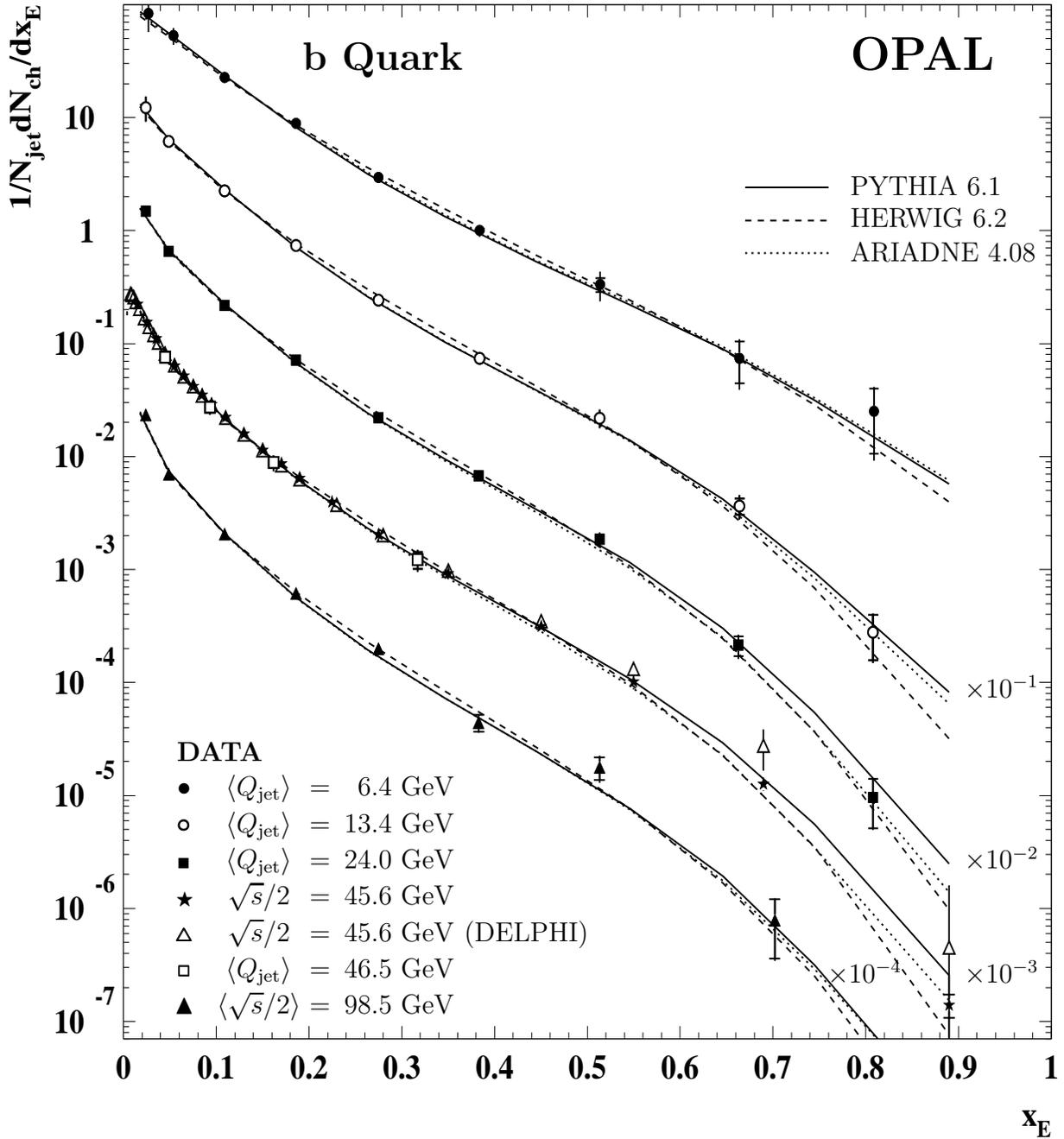,bbllx=68pt,bblly=180pt,bburx=525pt,%
bbury=680pt}
\caption{\xe\ dependence of the b jet \ffs\ at different scales 
compared with the PYTHIA 6.125, HERWIG 6.2 and ARIADNE 4.08 Monte Carlo 
predictions. The inner error bars indicate the statistical uncertainties, 
the total error bars show the statistical and systematic uncertainties added 
in quadrature. The values are given in Table~\ref{FFx}. For comparison, the 
published results on the unbiased jets of DELPHI \cite{B-decay} are shown.
The results are scaled as indicated in figure.}
\label{4scb-mod}
\end{figure}

\begin{figure}[tb] \centering
\epsfig{file=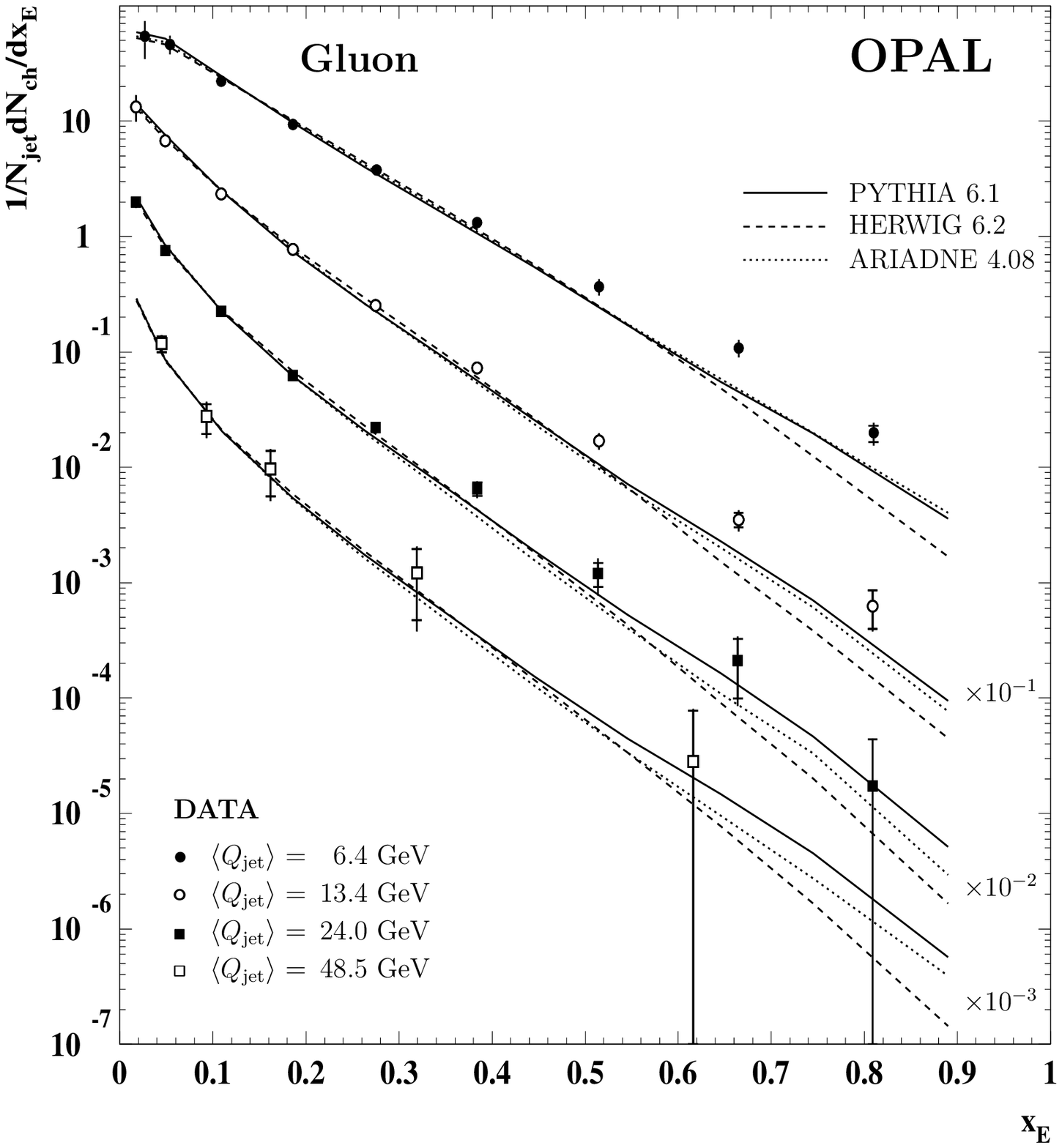,bbllx=68pt,bblly=180pt,bburx=525pt,%
bbury=680pt}
\caption{\xe\ dependence of the gluon jet \ffs\ at different scales 
compared with the PYTHIA 6.125, HERWIG 6.2 and ARIADNE 4.08 Monte Carlo 
predictions. The inner error bars indicate the statistical uncertainties,
the total error bars show the statistical and systematic uncertainties added 
in quadrature. The values are given in Table~\ref{FFx}.
The results are scaled as indicated in figure.}
\label{4scg-mod}
\end{figure}

In Figs.~\ref{4scl-nlo}--\ref{4scg-mod} we present the results shown in 
Figs.~\ref{fragl}--\ref{fragg} but now in a finer \xe\ binning and with the
additional data from the region $0.01<x_{\mr E}<0.03$, see Table \ref{FFx}. 
We integrate over the \ffs\ in four or five scale intervals:
$Q_{\mr jet}=$ 4--9, 9--19, 19--30, 30--70 and $\sqrt s/2=$ 91.5--104.5~GeV. 
Reference values for these intervals, evaluated as explained at the beginning 
of this section, are 6.4, 13.4, 24.0, 46.5 (48.5 for gluons) and 98.5 GeV, 
respectively. In the lowest scale interval, the data in the region of 
$0.01<x_{\mr E}<0.02$ are not measured due to the large dependence on the jet
algorithm.\\

In Figs.~\ref{4scl-nlo}--\ref{4scg-nlo} the data are compared 
to the NLO predictions. In general, the theory predictions are in a good 
agreement with the measurements of the udsc jet \ff\ (Fig.~\ref{4scl-nlo}).
We observe that the data in the region of low \xe\ are overestimated by the 
predictions of KKP, while they are in agreement with those of Kr and BFGW. 
For high \xe, the data prefer the KKP predictions but the differences between 
the predictions decrease with increasing scale.
In Fig.~\ref{4scb-nlo} the measured b jet \ff\ is shown together with the 
published results from DELPHI \cite{B-decay} and TPC \cite{TPC2}. Analogously 
to Fig.~\ref{fragb}, the spread of the NLO predictions is larger than that
for the udsc jet \ffs. The NLO predictions by Kr are seen to provide 
a reasonable description of all the b-jet data, while those by KKP and BFGW 
generally overestimate the data in the region of low \xe\ and underestimate 
them for large \xe. A possible explanation for this difference is that,
unlike KKP or BFGW, the fitting procedure of Kr includes both the low 
\xe\ (down to $x_{\mr E}=0.05$) and low scale data (TPC data \cite{TPC2} taken
at $\sqrt s=$29~GeV). In Fig.~\ref{4scg-nlo} the measured gluon jet \ffs\ are 
shown along with the OPAL \cite{indgl2} measurement at $E_{\mr jet} =$ 
14.24~GeV. An overall agreement is found between the results of the boost
method and the method used here. The observed sizable spread of the NLO 
predictions is expected because of the different approaches to the 
fitting procedures of the gluon jet data (see Section \ref{NLO}).\\

To test various fragmentation models, the data are also compared in 
Figs.~\ref{4scl-mod}--\ref{4scg-mod} to the hadron level predictions of the 
PYTHIA 6.125, HERWIG 6.2 and ARIADNE 4.08 MC event generators.
The hadron level is defined in Section \ref{correction}. 
Globally, all MC models give a more satisfactory description of the data than 
do the NLO predictions. This is presumably due to the fact that the biased jet 
data are compared to the biased jet MC predictions and the unbiased jet data 
to the unbiased jet MC predictions. We note that although all MC models used 
in this study were previously tuned to LEP1 data, they still provide a good 
description of the LEP2 data. There exist some discrepancies in the 
description of the gluon 
jet data in the region of high \xe\ with small scales (Fig.~\ref{4scg-mod}). 
A good agreement is achieved for the b jet \ffs\ by all three models. 

\subsection{Charged particle multiplicities}
By integrating the unbiased jet \ffs, the charged particle multiplicities in 
udsc, b and inclusive hadronic events can be obtained. The results for the 
LEP2 data are presented in the $\sqrt s$ intervals specified above, namely 
183--189, 192--202 and 204--209~GeV for the inclusive hadronic and 
udsc events and 183--209~GeV for the b events. 

\begin{tabbing}
aaaaaaaaaaaaa \= aaaaaaaaaaaaaaaaaaaaaaaaa \= aaaaaaaaaaaaaaaaaaaaaaaa \= \kill
$\ \ \ \ \ \langle \sqrt s \rangle$
\>$\ \ \ \ \ \ \ \langle n_{\mr ch}^{\mr incl} \rangle$ 
\>$\ \ \ \ \ \ \ \langle n_{\mr ch}^{\mr udsc} \rangle$  
\>$\ \ \ \ \ \ \ \langle n_{\mr ch}^{\mr b}    \rangle$  \\ 
$\ \,91.2$ GeV:\> $20.93 \pm 0.01 \pm 0.23$
               \> $20.32 \pm 0.03 \pm 0.27$
               \> $23.28 \pm 0.09 \pm 0.70$        \\
187.6 GeV:     \> $26.80 \pm 0.24 \pm 0.46$
               \> $26.43 \pm 0.26 \pm 0.81$        \\
197.0 GeV:     \>    
               \> 
               \> $30.01 \pm 0.53 \pm 0.82$        \\
198.0 GeV:     \> $27.68 \pm 0.26 \pm 0.50$   
               \> $27.38 \pm 0.31 \pm 0.85$
               \>                                  \\
206.2 GeV:     \> $27.75 \pm 0.29 \pm 0.67$
               \> $26.87 \pm 0.32 \pm 0.99$              
\end{tabbing}
The results are found to be in agreement with the previous unbiased jet 
measurements \cite{flavourff,B-decay,nch}. We also observed a good agreement
between the data and predictions of the three MC models used in this analysis.
The results for $\langle n_{\mr ch}^{\mr udsc} \rangle$ above 91.2~GeV energy 
represent new measurements. 

\section{Conclusions}\label{conclusions}
Scaling violations of quark and gluon jet \ffs\ are studied in \ee\ 
annihilations at $\sqrt s =$ 91.2 and 183--209~GeV using data collected 
with the OPAL detector at LEP. The scale dependence of the flavour inclusive, 
udsc and b \ffs\ from unbiased jets is measured at $\sqrt s/2=$ 45.6 and 
91.5--104.5~GeV. Biased jets are used to extract the flavour inclusive, udsc 
and b, and gluon \ffs\ in the ranges $Q_{\mr jet}=$ 4--42, 4--105 and 
4--70~GeV, respectively, where \Qj\ is the jet energy scale. Three methods 
are used to extract the \ffs, namely the b-tag and energy-ordering methods 
for biased jets, and the hemisphere method for unbiased jets. The results 
obtained using these methods are found to be consistent with each other. 
The udsc jet results above the scale of 45.6~GeV, the gluon jet results above 
30~GeV (except for the scale of 40.1~GeV), and the b jet results at all scales
except 45.6~GeV represent new measurements. The results of this analysis are 
compared with existing lower energy \ee\ data and with previous 
results from DELPHI and OPAL. The overall consistency of the biased jet 
results with the unbiased jet results suggests that \Qj\ is a generally 
appropriate scale in events with a general three-jet topology.  
The scaling violation is observed to be positive for lower \xe\ and 
negative for higher \xe, for all the types of \ffs. The gluon jet \ff\ exhibits
stronger scaling violation than that of udsc jets.\\

The bias of the procedure used to construct biased jet \ffs\ is estimated by
studying hadron level Monte Carlo generator events. In explaining the 
observed differences between biased and unbiased jet results, we note the 
effects of non-negligible masses of hadrons and b-quarks at low scales. 
Due to the considerable bias found for the gluon jet \ffs\ in the region of 
$x_{\mr E} > 0.6$, precautions should be taken when comparing the biased gluon 
jet results with theory.\\

The data are compared to the predictions of NLO calculations. In a wide 
range of \xe, all calculations satisfactorily describe the 
data for the udsc jet \ffs. 
The description is worse and the spread between the predictions larger for the 
b and gluon jet \ffs, in particular in regions of very low and high \xe. \\

The data are also compared with predictions of three Monte Carlo models, 
PYTHIA 6.125, HERWIG 6.2 and ARIADNE 4.08. 
A reasonable agreement with data is observed for all models, except for high 
\xe\ region with small scales ($\lesssim$ 14~GeV) in case of the udsc 
and gluon jet \ffs.\\

The charged particle multiplicities of udsc, b and inclusive hadronic 
events are obtained by integrating the measured \ffs. All values are 
found to be in agreement with previous measurements, where available.


\appendix
\par
\section*{Acknowledgements}
\par
We thank B. Kniehl, B. P\"{o}tter, K. Kramer and S. Kretzer for providing us
with their codes and for helpful discussions and J. Ch\'{y}la for valuable 
communication. \\   

We particularly wish to thank the SL Division for the efficient operation
of the LEP accelerator at all energies and for their close cooperation with
our experimental group.  In addition to the support staff at our own
institutions we are pleased to acknowledge the  \\
Department of Energy, USA, \\
National Science Foundation, USA, \\
Particle Physics and Astronomy Research Council, UK, \\
Natural Sciences and Engineering Research Council, Canada, \\
Israel Science Foundation, administered by the Israel
Academy of Science and Humanities, \\
Benoziyo Center for High Energy Physics,\\
Japanese Ministry of Education, Culture, Sports, Science and
Technology (MEXT) and a grant under the MEXT International
Science Research Program,\\
Japanese Society for the Promotion of Science (JSPS),\\
German Israeli Bi-national Science Foundation (GIF), \\
Bundesministerium f\"ur Bildung und Forschung, Germany, \\
National Research Council of Canada, \\
Hungarian Foundation for Scientific Research, OTKA T-038240, 
and T-042864,\\
The NWO/NATO Fund for Scientific Research, the Netherlands.\\


\vspace*{0.8cm}
\begin{table}[h]
\begin{center}
\begin{tabular}{||c||c|c|c||} 
\hhline{|t:====:t|}
Selection        &Data LEP1 &Data LEP2     &BG(LEP2)\\ \cline{1-4} \hline\hline
Hadronic events  &2 387 227 &10 866 (12 653)& 11\% (14\%) \\ \hline  
three-jet events &  965 513 & 6 177        & 16\% \\ \hline
udsc jets        &2 675 679 &16 344        & 16\% \\ \hline
b-tag jets       &   83 549 &   820        &  9\% \\ \hline
Gluon jets       &   73 620 &   729        &  9\% \\ \hline   
udsc hemispheres &4 740 774 & 20 146       & 11\% \\ \hline
b-tag hemispheres&   33 680 &  1 586       &  5\% \\  
\hhline{|b:====:b|}
\end{tabular}
\end{center}
\caption{Statistics of the LEP1 and LEP2 data samples. The number of hadronic 
events is given (the numbers in brackets correspond to LEP2 events used for 
three-jet analysis) and the characteristics of the biased jet samples (number 
of three-jet events, udsc, b-tag and gluon jets) and the unbiased jet samples 
(number of udsc and b-tag hemispheres) are shown. The jets are found by the 
Durham jet algorithm. Also indicated is the percentage of the remaining 
4-fermion background (BG) for the LEP2 data. For the LEP1 data, the background
is negligible.}
\label{LEP12stat} 
\end{table}

\vspace*{-3.5cm}
\begin{table}[th]
\begin{center}
\begin{tabular}{||c||c|c||} 
\hhline{|t:===:t|}
Cuts         &\multicolumn{2}{|c||}{\small Loss [\%]} \\ \cline{2-3}
             &\small LEP1 & \small LEP2 \\ \hline \hline 
Particle multiplicity per jet $\ge$ 2     &     0.7 & 1.7      \\ \hline 
Sum of inter-jet angles $\ge 358^\circ$   &     3.9 & 2.3      \\ \hline
Polar jet angle $|\cos \theta_{\rm jet}|\le 0.90\ (0.95)$&8.4&2.3  \\ \hline   
Corrected jet energy $\ge$ 5 GeV          &    11.2 & 5.9      \\ \hline   
Inter-jet angle $\ge 30^\circ$            &    43.3 & 43.2     \\ 
\hhline{|b:===:b|}
\end{tabular} 
\end{center}
\caption {Jet selection cuts for the LEP1 and LEP2 data and the reduction of 
statistics found by imposing each cut individually after the hadronic event 
selection. The jets are found by the Durham jet algorithm. The cut value given
in brackets corresponds to the LEP2 selection.}
\label{Jetcuts}
\end{table}

{\footnotesize
\begin{table}[ht]
\begin{center}
\begin{tabular}{||c|r@{\ \ -- \ }l||lll||c|r@{\ \ -- \ }l||lll||} 
\hhline{|t:============:t|}
\multicolumn{1}{||c|}{\x}&\multicolumn{2}{|c||}{\sa}&
\multicolumn{3}{c||}{\FFl}&
\multicolumn{1}{|c|}{\x}&\multicolumn{2}{|c||}{\sa}&
\multicolumn{3}{|c||}{\FFl}\\ \hline\hline
0.03--0.07&4.0& 6.5&38.1&\p1.5&\p4.0&0.22--0.48& 4.0& 6.5&2.54&\p0.14&\p0.17\\ 
        &  6.5& 9.0&45.5&\p1.1&\p4.7&      & 6.5& 9.0&2.28 &\p0.07 &\p0.15 \\
        &  9.0&12.0&44.8&\p0.7&\p2.3&      & 9.0&12.0&2.383&\p0.037&\p0.063\\ 
        & 12.0&15.0&49.8&\p0.7&\p2.6&      &12.0&15.0&2.205&\p0.032&\p0.059\\ 
        & 15.0&19.0&51.9&\p0.6&\p2.7&      &15.0&19.0&2.142&\p0.027&\p0.057\\ 
        & 19.0&24.0&54.12&\p0.55&\p0.94&   &19.0&24.0&2.074&\p0.024&\p0.026\\
        & 24.0&30.0&57.31&\p0.51&\p0.99&   &24.0&30.0&2.017&\p0.022&\p0.025\\ 
        & 30.0&42.0&55.6&\p0.2&\p2.2 &     &30.0&42.0&2.058&\p0.011&\p0.099\\ 
        & 30.0&47.0&49.3&\p4.9&\p3.8 &     &30.0&47.0&2.05 &\p0.21 &\p0.17 \\ 
        &\multicolumn{2}{c||}{45.6}&61.80&\p0.08 &\p0.82&
        &\multicolumn{2}{c||}{45.6}&1.899&\p0.004&\p0.038\\ 
        & 47.0& 70.0&60.6&\p3.2&\p4.7&     &47.0& 70.0&1.68&\p0.25&\p0.14 \\ 
        & 70.0&105.0&61.1&\p5.6&\p9.5&     &70.0&105.0&1.93&\p0.28&\p0.33 \\ 
        &\multicolumn{2}{c||}{93.8} &64.1  &\p0.8  &\p2.1&
        &\multicolumn{2}{c||}{93.8} &1.724 &\p0.040&\p0.040\\
        &\multicolumn{2}{c||}{99.0} &65.3  &\p0.9  &\p2.6&
        &\multicolumn{2}{c||}{99.0} &1.629 &\p0.047&\p0.042\\
        &\multicolumn{2}{c||}{103.1}&64.3  &\p1.0  &\p2.1&
        &\multicolumn{2}{c||}{103.1}&1.695 &\p0.050&\p0.037\\ 
\hline
0.07--0.12&4.0&6.5&20.9&\p1.0&\p1.0 &0.48--0.90&4.0&6.5&0.214&\p0.031&\p0.026\\
          & 6.5& 9.0&23.9 &\p0.6 &\p1.1 &    & 6.5 &9.0&0.219&\p0.013&\p0.026\\
          & 9.0&12.0&22.46&\p0.37&\p0.57&    & 9.0&12.0&0.201&\p0.006&\p0.012\\
          &12.0&15.0&22.55&\p0.35&\p0.57&    &12.0&15.0&0.183&\p0.006&\p0.011\\
          &15.0&19.0&22.88&\p0.31&\p0.58&&15.0&19.0&0.180&\p0.005&\p0.011\\ 
          &19.0&24.0&23.16&\p0.27&\p0.74&&19.0&24.0&0.1697&\p0.0040&\p0.0076\\ 
          &24.0&30.0&22.61&\p0.26&\p0.72&&24.0&30.0&0.1581&\p0.0036&\p0.0071\\ 
          &30.0&42.0&22.60&\p0.12&\p0.91&&30.0&42.0&0.1633&\p0.0015&\p0.0043\\ 
          &30.0&47.0&23.4 &\p2.2 &\p1.4 &&    &    &      &   --   &   \\ 
          &\multicolumn{2}{c||}{45.6}&23.57 &\p0.04  &\p0.24&    
          &\multicolumn{2}{c||}{45.6}&0.1411&\p0.0005&\p0.0058\\
          &47.0& 70.0&23.3&\p2.0 &\p1.4 &&30.0&70.0 &0.128&\p0.032&\p0.013\\
          &70.0&105.0&21.3&\p2.7 &\p3.3 &&70.0&105.0&0.052&\p0.073&\p0.019\\ 
          &\multicolumn{2}{c||}{93.8} &23.10 &\p0.41  &\p0.64&  
          &\multicolumn{2}{c||}{93.8} &0.1231&\p0.0079&\p0.0060\\
          &\multicolumn{2}{c||}{99.0} &23.89 &\p0.46  &\p0.99& 
          &\multicolumn{2}{c||}{99.0} &0.1154&\p0.0091&\p0.0060\\
          &\multicolumn{2}{c||}{103.1}&23.37 &\p0.52  &\p0.75&     
          &\multicolumn{2}{c||}{103.1}&0.1289&\p0.0095&\p0.0062\\
\hline
0.12--0.22&4.0& 6.5&10.25&\p0.47&\p0.69&        \multicolumn{6}{c}{}\\
        &  6.5& 9.0& 9.81&\p0.26&\p0.66&        \multicolumn{6}{c}{}\\
        &  9.0&12.0& 9.81&\p0.15&\p0.35&        \multicolumn{6}{c}{}\\
        & 12.0&15.0& 9.87&\p0.14&\p0.35&        \multicolumn{6}{c}{}\\    
        & 15.0&19.0& 9.44&\p0.12&\p0.34&        \multicolumn{6}{c}{}\\     
        & 19.0&24.0& 9.38&\p0.10&\p0.14&        \multicolumn{6}{c}{}\\       
        & 24.0&30.0& 9.36&\p0.09&\p0.14&        \multicolumn{6}{c}{}\\    
        & 30.0&42.0& 9.23&\p0.05&\p0.25&        \multicolumn{6}{c}{}\\    
        & 30.0&47.0& 9.60&\p0.81&\p0.58&        \multicolumn{6}{c}{}\\        
        &\multicolumn{2}{c||}{45.6}&8.98&\p0.02&\p0.14&\multicolumn{6}{c}{}\\  
        & 47.0&70.0& 8.24&\p0.89&\p0.49&        \multicolumn{6}{c}{}\\     
        & 70.0&105.0& 7.7&\p1.1 &\p0.8 &        \multicolumn{6}{c}{}\\      
        &\multicolumn{2}{c||}{93.8}&8.92&\p0.17&\p0.16&\multicolumn{6}{c}{}\\  
        &\multicolumn{2}{c||}{99.0}&8.48&\p0.19&\p0.29&\multicolumn{6}{c}{}\\ 
        &\multicolumn{2}{c||}{103.1}&8.68&\p0.20&\p0.08&\multicolumn{6}{c}{}\\
\hhline{|b:======:b|}
\end{tabular}
\end{center}
\caption{The udsc jet \ff\ in bins of \x\ and scale. The scale denotes 
\Qj\ for the biased jets and is given by the intervals, while it 
denotes $\sqrt s/2$ for the unbiased jets and is given by the single values. 
The first uncertainty is statistical, the second systematic. These data are 
displayed in Fig.~\ref{fragl}.}
\label{FFscl} 
\end{table}
}

{\footnotesize
\begin{table}[ht]
\begin{center}
\begin{tabular}{||c|r@{\ \ -- \ }l||lll||c|r@{\ \ -- \ }l||lll||} 
\hhline{|t:============:t|}
\multicolumn{1}{||c|}{\x}&\multicolumn{2}{|c||}{\sa}&
\multicolumn{3}{c||}{\FFb}&
\multicolumn{1}{|c|}{\x}&\multicolumn{2}{|c||}{\sa}&
\multicolumn{3}{|c||}{\FFb}\\ \hline\hline
0.03--0.07&4.0& 6.5&51.2&\p2.1&\p8.9&0.22--0.48& 4.0& 6.5&2.10&\p0.15&\p0.20\\ 
          & 6.5& 9.0&61  &\p1  &\p11 &      & 6.5& 9.0&1.74 &\p0.06 &\p0.16 \\
          & 9.0&12.0&65.6&\p0.8&\p4.5&      & 9.0&12.0&1.53 &\p0.04 &\p0.13 \\ 
          &12.0&15.0&65.7&\p0.7&\p4.5&      &12.0&15.0&1.45 &\p0.04 &\p0.12 \\ 
          &15.0&19.0&69.9&\p0.7&\p4.8&      &15.0&19.0&1.43 &\p0.03 &\p0.12 \\ 
          &19.0&24.0&70.8&\p0.6&\p2.9&      &19.0&24.0&1.360&\p0.029&\p0.055\\
          &24.0&30.0&72.2&\p0.6&\p3.0&      &24.0&30.0&1.359&\p0.026&\p0.055\\ 
          &30.0&42.0&71.4&\p0.3&\p3.3&      &30.0&42.0&1.373&\p0.015&\p0.022\\ 
          &30.0&47.0&72.9&\p6.4&\p5.1&      &30.0&47.0&1.37 &\p0.31 &\p0.19 \\ 
          &\multicolumn{2}{c||}{45.6}&76.9 &\p0.2  &\p1.9&
          &\multicolumn{2}{c||}{45.6}&1.253&\p0.011&\p0.064\\ 
          &47.0&70.0&77.0&\p6.2&\p5.4&       &47.0&70.0&1.12&\p0.29 &\p0.15 \\ 
          &70.0&105.0&74.3&\p4.8&\p7.3&      &70.0&105.0&1.34&\p0.21&\p0.18 \\ 
          &\multicolumn{2}{c||}{98.5}&77.4 &\p1.7  &\p1.7&
          &\multicolumn{2}{c||}{98.5}&1.112&\p0.075&\p0.089\\
\hline
0.07--0.12& 4.0&6.5&28.7&\p1.3&\p1.5 &0.48--0.90& 4.0&6.5&0.134&\p0.046&\p0.034\\
        & 6.5&9.0 &30.1 &\p0.7 &\p1.6 &  & 6.5&9.0 &0.111 &\p0.020 &\p0.028 \\
        & 9.0&12.0&30.32&\p0.42&\p0.48&  & 9.0&12.0&0.076 &\p0.010 &\p0.013 \\
        &12.0&15.0&30.77&\p0.41&\p0.49&  &12.0&15.0&0.066 &\p0.008 &\p0.012 \\
        &15.0&19.0&30.21&\p0.35&\p0.48&  &15.0&19.0&0.0564&\p0.0067&\p0.0099\\ 
        &19.0&24.0&29.96&\p0.32&\p0.63&  &19.0&24.0&0.0466&\p0.0053&\p0.0066\\ 
        &24.0&30.0&29.90&\p0.29&\p0.62&  &24.0&30.0&0.0524&\p0.0052&\p0.0074\\ 
        &30.0&42.0&29.71&\p0.17&\p0.60&  &30.0&42.0&0.0476&\p0.0029&\p0.0068\\ 
        &30.0&47.0&28.6 &\p3.4 &\p3.5 &  &30.0&47.0&      & --     &\\ 
        &\multicolumn{2}{c||}{45.6}&30.15&\p0.12   &\p0.44&    
        &\multicolumn{2}{c||}{45.6}&0.0379&\p0.0016&\p0.0041\\
        &47.0&70.0&26.5 &\p3.1 &\p3.3 &  &47.0&70.0&      & --     &\\
        &70.0&105.0&30.9&\p2.5 &\p3.9 &&70.0&105.0&0.056&\p0.040&\p0.016    \\ 
    &\multicolumn{2}{c||}{98.5}&28.9&\p0.9&\p1.4&  
    &\multicolumn{2}{c||}{98.5}&0.046&\p0.011&\p0.007\\
\hline
0.12--0.22&4.0&6.5&11.80&\p0.54&\p0.75&         \multicolumn{6}{c}{}\\
        & 6.5&9.0 &11.14&\p0.25&\p0.71&         \multicolumn{6}{c}{}\\
        & 9.0&12.0&10.41&\p0.15&\p0.47&         \multicolumn{6}{c}{}\\
        &12.0&15.0& 9.94&\p0.15&\p0.45&         \multicolumn{6}{c}{}\\    
        &15.0&19.0& 9.99&\p0.13&\p0.45&         \multicolumn{6}{c}{}\\     
        &19.0&24.0& 9.98&\p0.12&\p0.28&         \multicolumn{6}{c}{}\\       
        &24.0&30.0& 9.58&\p0.11&\p0.26&         \multicolumn{6}{c}{}\\    
        &30.0&42.0& 9.83&\p0.06&\p0.27&         \multicolumn{6}{c}{}\\    
        &30.0&47.0& 8.7 &\p1.2 &\p1.2 &         \multicolumn{6}{c}{}\\        
    &\multicolumn{2}{c||}{45.6}&9.40&\p0.05&\p0.28&\multicolumn{6}{c}{}\\   
        &47.0&70.0& 9.0&\p1.1&\p1.2&            \multicolumn{6}{c}{}\\     
        &70.0&105.0&8.8&\p0.9 &\p1.0 &          \multicolumn{6}{c}{}\\      
    &\multicolumn{2}{c||}{98.5}&8.94&\p0.33&\p0.24&\multicolumn{6}{c}{}\\  
\hhline{|b:======:b|}
\end{tabular}
\end{center}
\caption{The b jet \ff\ in bins of \xe\ and scale. The scale denotes 
\Qj\ for the biased jets and is given by the intervals, while it 
denotes $\sqrt s/2$ for the unbiased jets and is given by the single values.
The first uncertainty is statistical, the second systematic. These data are 
displayed in Fig.~\ref{fragb}. In the region 0.48 $<x_{\mr E}<$ 0.90 and 
$Q_{\mr jet}=$ 30--70~GeV, no measurement was possible due to low  statistics.}
\label{FFscb} 
\end{table}
}

{\footnotesize
\begin{table}[ht]
\begin{center}
\begin{tabular}{||c|r@{\ \ -- \ }l||lll||r@{\ \  -- \ }l||lll||} 
\hhline{|t:===========:t|}
\multicolumn{1}{||c|}{\x}&\multicolumn{2}{|c||}{\Q}&
\multicolumn{3}{c||}{\FFg\ (BT)}&
\multicolumn{2}{|c||}{\Q}&\multicolumn{3}{|c||}{\FFg\ (EO)}\\ \hline\hline
0.03--0.07&4.0& 6.5&43.3&\p0.6&\p9.6   & 6.0& 6.5&44&\p1&\p10   \\   
        & 6.5 & 9.0&58.2&\p0.6&\p9.4   & 6.5& 9.0&57.1&\p0.4&\p6.4    \\  
        & 9.0 &12.0&69.9&\p0.8&\p6.4   & 9.0&12.0&68.1&\p0.4&\p5.8    \\  
        &12.0 &15.0&73.8&\p1.0&\p6.8   &12.0&15.0&75.7&\p0.6&\p4.4    \\ 
        &15.0 &19.0&79.5&\p1.3&\p7.3   &15.0&19.0&80.6&\p0.8&\p4.4    \\ 
        &19.0 &24.0&84.7&\p1.7&\p6.8   &19.0&24.0&85.6&\p1.4&\p5.2    \\ 
        &24.0 &30.0&80.5&\p2.4&\p6.4   &24.0&27.0&92.4&\p5.3&\p5.6    \\ 
        &30.0 &42.0&89  &\p4  &\p14    &    &    &    & --  &         \\
        &30.0 &70.0&118 &\p19 &\p10    &30.0&60.0&101&\p11&\p9    \\ 
\hline
0.07--0.12&4.0&6.5 &26.6 &\p0.3 &\p1.4 & 6.0&6.5 &27.5 &\p0.5 &\p2.3    \\    
        & 6.5 &9.0 &29.7 &\p0.4 &\p1.6 & 6.5&9.0 &30.4 &\p0.2 &\p1.0 \\ 
        & 9.0 &12.0&31.65&\p0.43&\p0.71& 9.0&12.0&32.37&\p0.21&\p0.86 \\ 
        &12.0 &15.0&32.69&\p0.58&\p0.74&12.0&15.0&33.00&\p0.30&\p0.81 \\ 
        &15.0 &19.0&32.14&\p0.66&\p0.73&15.0&19.0&32.90&\p0.40&\p0.87 \\ 
        &19.0 &24.0&31.2 &\p0.9 &\p1.8 &19.0&24.0&32.2&\p0.7  &\p1.1 \\  
        &24.0 &30.0&33.6 &\p1.5 &\p1.9 &24.0&27.0&34.7 &\p2.7 &\p1.2  \\
        &30.0 &42.0&32.7 &\p2.1 &\p2.7 &    &    &     & --   &       \\ 
        &30.0 &70.0&27.5 &\p7.9 &\p5.2 &30.0&60.0&30.2 &\p5.0 &\p2.8  \\
\hline
0.12--0.22&4.0& 6.5&12.14&\p0.15&\p0.25& 6.0& 6.5&11.99&\p0.19&\p0.17  \\    
        & 6.5 & 9.0&11.88&\p0.16&\p0.24& 6.5& 9.0&12.10&\p0.08&\p0.21  \\ 
        & 9.0 &12.0&11.12&\p0.17&\p0.20& 9.0&12.0&11.49&\p0.09&\p0.29  \\ 
        &12.0 &15.0&10.39&\p0.22&\p0.19&12.0&15.0&10.87&\p0.12&\p0.35  \\ 
        &15.0 &19.0&10.01&\p0.26&\p0.18&15.0&19.0&10.35&\p0.16&\p0.26  \\ 
        &19.0 &24.0& 9.26&\p0.34&\p0.68&19.0&24.0& 9.54&\p0.26&\p0.35  \\ 
        &24.0 &30.0& 8.46&\p0.52&\p0.62&24.0&27.0& 8.13&\p0.86&\p0.30  \\
        &30.0 &42.0& 7.5 &\p0.7 &\p1.7 &    &    &     & --   &        \\
        &30.0 &70.0& 9.7 &\p4.1 &\p2.1 &30.0&60.0&12.6 &\p2.2 &\p2.3   \\
\hline
0.22--0.48&4.0& 6.5&2.603&\p0.041&\p0.092&6.0&6.5  &2.59 &\p0.05  &\p0.12 \\   
        & 6.5 & 9.0&2.022&\p0.038&\p0.072& 6.5&9.0 &2.034&\p0.020&\p0.052 \\ 
        & 9.0 &12.0&1.587&\p0.038&\p0.076& 9.0&12.0&1.741&\p0.021&\p0.071 \\ 
        &12.0 &15.0&1.527&\p0.052&\p0.073&12.0&15.0&1.589&\p0.031&\p0.049 \\ 
        &15.0 &19.0&1.403&\p0.062&\p0.067&15.0&19.0&1.389&\p0.039&\p0.090 \\ 
        &19.0 &24.0&1.33 &\p0.08 &\p0.13 &19.0&24.0&1.21 &\p0.07 &\p0.18  \\ 
        &24.0 &30.0&1.35 &\p0.14 &\p0.14 &24.0&27.0&1.14 &\p0.24 &\p0.17  \\
        &30.0 &42.0&1.22 &\p0.18 &\p0.13 &    &    &     &  --   &        \\ 
        &30.0 &70.0&1.21 &\p0.74 &\p0.38 &30.0&60.0&1.03 &\p0.49 &\p0.15  \\
\hline
0.48--0.90&4.0& 6.5&0.168 &\p0.008&\p0.026  & 6.0& 6.5&0.158&\p0.010&\p0.023\\ 
        &  6.5& 9.0&0.085 &\p0.006&\p0.013  & 6.5& 9.0&0.099&\p0.005&\p0.014\\ 
        & 9.0 &12.0&0.069 &\p0.007 &\p0.010 & 9.0&12.0&0.077&\p0.006&\p0.016\\ 
        &12.0 &15.0&0.0527&\p0.0072&\p0.0078&12.0&15.0&0.072&\p0.013&\p0.023\\ 
        &15.0 &19.0&0.0350&\p0.0066&\p0.0052&15.0&19.0&0.047&\p0.014&\p0.015\\ 
        &19.0 &24.0&0.033 &\p0.009 &\p0.009 &19.0&24.0&0.019&\p0.047&\p0.006\\
        &24.0 &30.0&0.064 &\p0.033 &\p0.017 &24.0&27.0&     &    -- &       \\
        &30.0 &42.0&      &   --   &        &    &    &     &    -- &       \\ 
        &30.0 &70.0&0.028 &\p0.050 &\p0.017 &30.0&60.0&     &    -- &       \\ 
\hhline{|b:===========:b|}
\end{tabular}
\end{center}
\caption{The gluon jet \ffs\ in bins of \xe\ and scale \Qj\ obtained 
from the biased jets using the b-tag method (BT) and the energy-ordering method
(EO). The first uncertainty is statistical, the second systematic. These data 
are displayed in Fig.~\ref{fragg}. In the region 0.48 $<x_{\mr E}<$ 0.90 and 
$Q_{\mr jet}=$ 30--42~GeV for the b-tag method and $Q_{\mr jet}=$ 24--60~GeV 
for the energy-ordering method, no measurement was possible due to low 
statistics.}
\label{FFscg} 
\end{table}
}

{\scriptsize
\begin{table}[ht]
\begin{center}
\begin{tabular}{||c|r@{\ \ -- \ }l||lll||c|r@{\ \ -- \ }l||lll||} 
\hhline{|t:============:t|}
\multicolumn{1}{||c|}{\x}&\multicolumn{2}{|c||}{\sa}&
\multicolumn{3}{|c||}{\FFi}&
\multicolumn{1}{|c|}{\x}&\multicolumn{2}{|c||}{\sa}&
\multicolumn{3}{|c||}{\FFi}\\
\cline{1-4}\hline\hline
0.02--0.04& 4.0& 6.5 &46.0&\p3.4&\p9.3 &0.20--0.30& 4.0& 6.5&4.57&\p0.46&\p0.40\\
        & 6.5& 9.0 &69  &\p2  &\p14  &     & 6.5& 9.0&4.55 &\p0.20 &\p0.40\\
        & 9.0&12.0 &68.9&\p1.2&\p7.8 &     & 9.0&12.0&4.58 &\p0.09 &\p0.11\\
        &12.0&15.0 &75.5&\p1.2&\p8.5 &     &12.0&15.0&4.33 &\p0.08 &\p0.11\\
        &15.0&19.0 &80.5&\p1.1&\p9.1 &     &15.0&19.0&4.23 &\p0.07 &\p0.10\\
        &19.0&24.0 &87.7&\p1.0&\p2.1 &     &19.0&24.0&4.098&\p0.059&\p0.065\\
        &24.0&30.0 &94.8&\p1.0&\p2.3 &     &24.0&30.0&3.979&\p0.052&\p0.063\\
        &30.0&42.0 &98.7&\p0.5&\p4.6 &     &30.0&42.0&4.005&\p0.025&\p0.098\\
        &\multicolumn{2}{c||}{45.6}  &115.5&\p0.1    &\p1.7  &      
        &\multicolumn{2}{c||}{45.6}  &3.704&\p0.005  &\p0.029\\
        &\multicolumn{2}{c||}{93.8}  &123.9&\p1.2    &\p2.1  &        
        &\multicolumn{2}{c||}{93.8}  &3.525&\p0.076  &\p0.065\\ 
        &\multicolumn{2}{c||}{99.0}  &124.4&\p1.4    &\p2.9  &        
        &\multicolumn{2}{c||}{99.0}  &3.373&\p0.082  &\p0.076\\ 
        &\multicolumn{2}{c||}{103.1} &125.4&\p1.5    &\p2.3  &        
        &\multicolumn{2}{c||}{103.1} &3.372&\p0.089  &\p0.083\\ 
\hline
0.04--0.06& 4.0& 6.5 &41.5&\p4.1&\p3.0 &0.30--0.40& 4.0& 6.5&2.23&\p0.31&\p0.05\\
        & 6.5& 9.0 &45.0&\p2.0  &\p3.3 &  & 6.5& 9.0&1.72 &\p0.11 &\p0.04 \\
        & 9.0&12.0 &44.9&\p1.0  &\p2.1 &  & 9.0&12.0&1.994&\p0.055&\p0.066\\
        &12.0&15.0 &49.4&\p0.9  &\p2.3 &  &12.0&15.0&1.744&\p0.046&\p0.058\\
        &15.0&19.0 &51.9&\p0.8  &\p2.4 &  &15.0&19.0&1.692&\p0.040&\p0.056\\
        &19.0&24.0 &53.38&\p0.72&\p0.75&  &19.0&24.0&1.645&\p0.033&\p0.039\\
        &24.0&30.0 &56.42&\p0.63&\p0.80&  &24.0&30.0&1.619&\p0.030&\p0.038\\
        &30.0&42.0 &55.1 &\p0.3 &\p1.5 &  &30.0&42.0&1.613&\p0.016&\p0.070\\
        &\multicolumn{2}{c||}{45.6}&61.42 &\p0.05   &\p0.53  &        
        &\multicolumn{2}{c||}{45.6}&1.428 &\p0.003  &\p0.012 \\
        &\multicolumn{2}{c||}{93.8}&62.4  &\p0.8    &\p1.1   &          
        &\multicolumn{2}{c||}{93.8}&1.259 &\p0.045  &\p0.060 \\
        &\multicolumn{2}{c||}{99.0}&62.9  &\p0.9    &\p1.4   &          
        &\multicolumn{2}{c||}{99.0}&1.223 &\p0.048  &\p0.048 \\
        &\multicolumn{2}{c||}{103.1}&62.8 &\p1.0    &\p0.9  &          
        &\multicolumn{2}{c||}{103.1}&1.292&\p0.054  &\p0.043 \\ 
\hline
0.06--0.10& 4.0& 6.5 &30.0&\p2.8&\p2.8 &0.40--0.60& 4.0& 6.5&0.664&\p0.099&\p0.046\\  
        & 6.5& 9.0 &31.6 &\p1.1 &\p3.0    &&6.5& 9.0&0.658&\p0.040&\p0.046\\
        & 9.0&12.0 &28.2&\p0.5&\p1.0&     & 9.0&12.0&0.625&\p0.018&\p0.014\\
        &12.0&15.0 &29.3&\p0.5&\p1.1&     &12.0&15.0&0.575&\p0.017&\p0.013\\
        &15.0&19.0 &30.2&\p0.4&\p1.1&     &15.0&19.0&0.564&\p0.014&\p0.013\\
        &19.0&24.0 &29.79&\p0.35&\p0.45&  &19.0&24.0&0.546&\p0.012&\p0.013\\
        &24.0&30.0 &30.11&\p0.32&\p0.45&  &24.0&30.0&0.507&\p0.010&\p0.012\\
        &30.0&42.0 &30.26&\p0.15&\p0.96&  &30.0&42.0&0.508&\p0.005&\p0.037\\
        &\multicolumn{2}{c||}{45.6}&32.32 &\p0.03   &\p0.19  &        
        &\multicolumn{2}{c||}{45.6}&0.4241&\p0.0013 &\p0.0067\\ 
        &\multicolumn{2}{c||}{93.8}&31.56 &\p0.38   &\p0.61  &           
        &\multicolumn{2}{c||}{93.8}&0.416 &\p0.018  &\p0.020 \\
        &\multicolumn{2}{c||}{99.0}&31.93 &\p0.43   &\p0.83  &           
        &\multicolumn{2}{c||}{99.0}&0.389 &\p0.018  &\p0.021 \\
        &\multicolumn{2}{c||}{103.1}&31.70&\p0.46   &\p0.59  &           
        &\multicolumn{2}{c||}{103.1}&0.403&\p0.020  &\p0.023 \\ 
\hline
0.10--0.14& 4.0& 6.5 &14.8 &\p1.6 &\p0.8 &0.60--0.80& 4.0& 6.5&0.102&\p0.029&\p0.010\\ 
 
        & 6.5& 9.0 &16.07&\p0.72&\p0.87& & 6.5& 9.0&0.137 &\p0.014 &\p0.013 \\ 
        & 9.0&12.0 &16.75&\p0.36&\p0.38& & 9.0&12.0&0.1251&\p0.0070&\p0.0016\\
        &12.0&15.0 &16.38&\p0.31&\p0.37& &12.0&15.0&0.1177&\p0.0057&\p0.0015\\
        &15.0&19.0 &16.23&\p0.27&\p0.37& &15.0&19.0&0.1146&\p0.0047&\p0.0015\\
        &19.0&24.0 &16.29&\p0.23&\p0.24& &19.0&24.0&0.1057&\p0.0041&\p0.0056\\
        &24.0&30.0 &15.85&\p0.20&\p0.24& &24.0&30.0&0.0958&\p0.0036&\p0.0050\\
        &30.0&42.0 &15.99&\p0.10&\p0.51& &30.0&42.0&0.0922&\p0.0015&\p0.0038\\
        &\multicolumn{2}{c||}{45.6} &16.508 &\p0.018  &\p0.055   &          
        &\multicolumn{2}{c||}{45.6} &0.0755&\p0.0005&\p0.0029 \\
        &\multicolumn{2}{c||}{93.8} &16.14 &\p0.26  &\p0.46   &          
        &\multicolumn{2}{c||}{93.8} &0.0591&\p0.0062&\p0.0024 \\
        &\multicolumn{2}{c||}{99.0} &16.14 &\p0.29  &\p0.59   &          
        &\multicolumn{2}{c||}{99.0} &0.0644&\p0.0072&\p0.0026 \\
        &\multicolumn{2}{c||}{103.1}&16.11 &\p0.31  &\p0.34   &         
        &\multicolumn{2}{c||}{103.1}&0.0655&\p0.0077&\p0.0027 \\ 
\hline
0.14--0.20& 4.0& 6.5&10.5&\p1.1&\p0.5&0.80--1.00& 4.0& 6.5 &0.049 &\p0.012 &\p0.010 \\ 
 
       & 6.5& 9.0& 9.39&\p0.41&\p0.49&  & 6.5& 9.0 &0.0245&\p0.0044&\p0.0052\\ 
       & 9.0&12.0& 9.46&\p0.20&\p0.25&  & 9.0&12.0 &0.0245&\p0.0022&\p0.0087\\
       &12.0&15.0& 9.65&\p0.17&\p0.25&  &12.0&15.0 &0.0181&\p0.0015&\p0.0064\\
       &15.0&19.0& 8.99&\p0.14&\p0.23&  &15.0&19.0 &0.0154&\p0.0014&\p0.0055\\
       &19.0&24.0& 9.14&\p0.13&\p0.28&  &19.0&24.0 &0.0138&\p0.0010&\p0.0051\\
       &24.0&30.0& 9.08&\p0.11&\p0.27&  &24.0&30.0 &0.0125&\p0.0010&\p0.0046\\
       &30.0&42.0& 8.95&\p0.06&\p0.13&  &30.0&42.0 &0.0119&\p0.0004&\p0.0054\\
       &\multicolumn{2}{c||}{45.6}&8.638 &\p0.011  &\p0.036  &      
       &\multicolumn{2}{c||}{45.6}&0.0097&\p0.0001&\p0.0042 \\
       &\multicolumn{2}{c||}{93.8}&8.30  &\p0.15   &\p0.10   &         
       &\multicolumn{2}{c||}{93.8}&0.0112&\p0.0018 &\p0.0065 \\
       &\multicolumn{2}{c||}{99.0}&8.12  &\p0.17   &\p0.20   &         
       &\multicolumn{2}{c||}{99.0}&0.0095&\p0.0019 &\p0.0055 \\
       &\multicolumn{2}{c||}{103.1}&8.45 &\p0.18   &\p0.19   &         
       &\multicolumn{2}{c||}{103.1}&0.0106&\p0.0022&\p0.0062 \\
\hhline{|b:============:b|}
\end{tabular}
\end{center}
\caption{\small The flavour inclusive jet \ffs\ in bins of \xe\ and
scale. The scale denotes \Qj\ for the biased jets and is given by 
the intervals, while it denotes $\sqrt s/2$ for the unbiased jets 
and is given by the single values. The first uncertainty is statistical, 
the second systematic. These data are displayed in Figs.~\ref{fraginc} and 
\ref{4sci-nlo}.}
\label{FFincl} 
\end{table}
}
\clearpage
{\scriptsize
\begin{table}[ht]
\begin{center}
\begin{tabular}{||c|r@{\ \ -- \ }l||lll||lll||lll||} 
\hhline{|t:============:t|}
\multicolumn{1}{||c|}{\sa}&\multicolumn{2}{|c||}{\x}&
\multicolumn{3}{|c||}{\FFl}&\multicolumn{3}{|c||}{\FFb}&
\multicolumn{3}{|c||}{\FFg}\\ \hline\hline
 4.0--9.0& 0.02&0.04&59    &\p1     &\p15    &84    &\p2     &\p26
                    &54    &\p1     &\p20  \\   
        & 0.04&0.08 &40.1  &\p0.8   &\p4.2   &53.2  &\p0.9   &\p9.2 
                    &46.3  &\p0.3   &\p8.4  \\  
        & 0.08&0.15 &17.61 &\p0.40  &\p0.80  &22.7  &\p0.4   &\p1.2 
                    &22.3  &\p0.2   &\p1.2   \\  
        & 0.15&0.23 &8.23  &\p0.24  &\p0.56  &8.90  &\p0.24  &\p0.57
                    &9.37  &\p0.10  &\p0.19  \\ 
        & 0.23&0.33 &3.58  &\p0.14  &\p0.24  &2.94  &\p0.13  &\p0.28
                    &3.76  &\p0.06  &\p0.13  \\ 
        & 0.33&0.45 &1.469 &\p0.081 &\p0.098 &1.010 &\p0.082 &\p0.095
                    &1.323 &\p0.030 &\p0.047 \\ 
        & 0.45&0.60 &0.558 &\p0.042 &\p0.067 &0.334 &\p0.048 &\p0.085
                    &0.367 &\p0.014 &\p0.057 \\ 
        & 0.60&0.75 &0.136 &\p0.018 &\p0.016 &0.074 &\p0.030 &\p0.019
                    &0.108 &\p0.008 &\p0.017 \\
        & 0.75&0.90 &0.0724&\p0.0096&\p0.0087&0.025 &\p0.015 &\p0.006
                    &0.0198&\p0.0032&\p0.0031\\ \hline
 9.0--19.0& 0.01&0.03&94    &\p1     &\p23    &123    &\p1      &\p31 
                     &133   &\p1     &\p34    \\   
        & 0.03&0.08  &44.9  &\p0.3   &\p2.3   &61.8   &\p0.4    &\p4.3  
                     &67.7  &\p0.5   &\p6.2   \\  
        & 0.08&0.15  &17.78 &\p0.14  &\p0.45  &22.51  &\p0.16   &\p0.27
                     &23.44 &\p0.22  &\p0.53  \\  
        & 0.15&0.23  & 7.74 &\p0.08  &\p0.28  & 7.41  &\p0.08   &\p0.34
                     &7.74  &\p0.12  &\p0.14  \\ 
        & 0.23&0.33  &3.401 &\p0.040 &\p0.090 & 2.42  &\p0.04   &\p0.20
                     &2.54  &\p0.06  &\p0.12  \\ 
        & 0.33&0.45  &1.372 &\p0.021 &\p0.036 &0.738  &\p0.024  &\p0.062
                     &0.727 &\p0.029 &\p0.035 \\ 
        & 0.45&0.60  &0.489 &\p0.010 &\p0.030 &0.219  &\p0.014  &\p0.039
                     &0.170 &\p0.013 &\p0.025 \\ 
        &0.60&0.75   &0.1501&\p0.0045&\p0.0091&0.0366 &\p0.0062 &\p0.0064
                     &0.0352&\p0.0051&\p0.0052\\
        &0.75&0.90   &0.0404&\p0.0019&\p0.0025&0.00277&\p0.00119&\p0.00049
                     &0.0063&\p0.0023&\p0.0009\\ \hline
19.0--30.0&0.01&0.03&120.2  &\p0.8   &\p6.2   &148.1  &\p1.0    &\p8.7
                    &200    &\p3     &\p23    \\   
        &0.03&0.08  &50.70  &\p0.31  &\p0.88  &65.5   &\p0.4    &\p2.7
                    &76.0   &\p1.2   &\p6.1   \\  
        &0.08&0.15  &17.64  &\p0.13  &\p0.56  &21.88  &\p0.15   &\p0.46
                    &22.4   &\p0.5   &\p1.3   \\  
        &0.15&0.23  &7.39   &\p0.06  &\p0.11  &7.12   &\p0.08   &\p0.20
                    & 6.23  &\p0.27  &\p0.46  \\ 
        &0.23&0.33  &3.140  &\p0.035 &\p0.039 &2.207  &\p0.041  &\p0.089
                    & 2.22  &\p0.15  &\p0.23  \\ 
        &0.33&0.45  &1.248  &\p0.017 &\p0.015 &0.675  &\p0.022  &\p0.027
                    &0.652  &\p0.087 &\p0.066 \\ 
        &0.45&0.60  &0.427  &\p0.008 &\p0.019 &0.185  &\p0.012  &\p0.026
                    &0.120  &\p0.028 &\p0.032 \\ 
        &0.60&0.75  &0.1296 &\p0.0035&\p0.0058&0.0214 &\p0.0043 &\p0.0030
                    &0.021  &\p0.011&\p0.006  \\
        &0.75&0.90  &0.0314 &\p0.0014&\p0.0014&0.00096&\p0.00045&\p0.00014
                    &0.0017 &\p0.0027&\p0.0005\\ \hline
30.0--70.0&0.03&0.07&54.9   &\p2.9   &\p4.3   &75.4 &\p4.5   &\p5.3
                    &118    &\p19    &\p10    \\   
        &0.07&0.12  &23.4   &\p1.4   &\p1.4   &27.4 &\p2.3   &\p3.4
                    &27.5   &\p7.9   &\p5.2   \\  
        &0.12&0.22  &8.98   &\p0.56  &\p0.54  &8.8  &\p0.8   &\p1.2
                    &9.7    &\p4.1   &\p2.1   \\  
        &0.22&0.48  &1.90   &\p0.15  &\p0.16  &1.22 &\p0.21  &\p0.16
                    & 1.21  &\p0.74  &\p0.38  \\ 
        &0.48&0.90  &0.128  &\p0.032 &\p0.013 &     & --     &
                    & 0.028 &\p0.050 &\p0.017 \\   \hline

45.6  & 0.00&0.01&172.4 &\p0.4   &\p3.1     &185.9  &\p1.2    &\p7.8    \\
      & 0.01&0.02&201.3 &\p0.4   &\p3.1     &224    &\p1      &\p11     \\
      & 0.02&0.03&131.6 &\p0.3   &\p2.1     &154.3  &\p0.9    &\p6.5    \\
      & 0.03&0.04&90.6  &\p0.2   &\p1.4     &110.2  &\p0.7    &\p3.4    \\
      & 0.04&0.05&66.16 &\p0.21  &\p0.83    &82.3   &\p0.6    &\p1.9    \\
      & 0.05&0.06&50.72 &\p0.19  &\p0.72    &63.7   &\p0.5    &\p1.6    \\
      & 0.06&0.07&39.89 &\p0.17  &\p0.51    &51.90  &\p0.44   &\p0.61   \\
      & 0.07&0.08&32.45 &\p0.15  &\p0.25    &42.05  &\p0.40   &\p0.43   \\
      & 0.08&0.09&26.78 &\p0.14  &\p0.26    &35.17  &\p0.37   &\p0.53   \\
      & 0.09&0.10&22.69 &\p0.13  &\p0.21    &28.99  &\p0.33   &\p0.44   \\
      & 0.10&0.12&17.99 &\p0.08  &\p0.19    &22.33  &\p0.20   &\p0.40   \\
      & 0.12&0.14&13.64 &\p0.07  &\p0.17    &15.86  &\p0.17   &\p0.36   \\
      & 0.14&0.16&10.66 &\p0.06  &\p0.16    &11.32  &\p0.15   &\p0.27   \\
      & 0.16&0.18&8.36  &\p0.05  &\p0.16    &8.63   &\p0.13   &\p0.33   \\
      & 0.18&0.20&6.75  &\p0.04  &\p0.15    &6.43   &\p0.11   &\p0.43   \\
      & 0.20&0.25&4.791 &\p0.020 &\p0.077   &3.97   &\p0.06   &\p0.15   \\
      & 0.25&0.30&3.002 &\p0.014 &\p0.062   &2.05   &\p0.04   &\p0.11   \\
      & 0.30&0.40&1.570 &\p0.007 &\p0.027   &0.921  &\p0.020  &\p0.039  \\
      & 0.40&0.50&0.675 &\p0.004 &\p0.014   &0.314  &\p0.012  &\p0.012  \\
      & 0.50&0.60&0.296 &\p0.002 &\p0.012   &0.102  &\p0.007  &\p0.013  \\
      & 0.60&0.80&0.0926&\p0.0007&\p0.0046  &0.0127 &\p0.0013 &\p0.0012 \\   
      & 0.80&1.00&0.0119&\p0.0002&\p0.0061  &0.00014&\p0.00003&\p0.00011\\ 
\cline{1-9}

91.5--104.5&0.01&0.03&207.3  &\p1.0   &\p8.0   &234.5  &\p2.8   &\p6.2    \\  
           &0.03&0.08&58.4   &\p0.5   &\p2.0   &70.3   &\p1.4   &\p1.5    \\
           &0.08&0.15&17.57  &\p0.19  &\p0.58  &20.7   &\p0.6   &\p1.0    \\
           &0.15&0.23& 6.69  &\p0.10  &\p0.12  & 6.18  &\p0.30  &\p0.19   \\
           &0.23&0.33&2.597  &\p0.055 &\p0.061 & 2.01  &\p0.17  &\p0.18   \\
           &0.33&0.45&1.002  &\p0.029 &\p0.029 &0.442  &\p0.076 &\p0.032  \\
           &0.45&0.60&0.326  &\p0.015 &\p0.015 &0.178  &\p0.041 &\p0.025  \\
           &0.60&0.90&0.0550 &\p0.0037&\p0.0032&0.0079 &\p0.0043&\p0.0012 \\
\hhline{|b:=========:b|}
\end{tabular}
\end{center}
\vspace*{-0.2cm}
\caption{\small The udsc-quark, b-quark and gluon \ffs\ in bins of \xe\ and 
scale. The scale denotes \Qj\ for the biased jets (the first four intervals) 
and $\sqrt s/2$ for the unbiased jets (the single value and the last 
interval). The first uncertainty is statistical, the second systematic. 
These data are displayed in Figs.~\ref{4scl-nlo}--\ref{4scg-mod}.}
\label{FFx} 
\end{table}
}


\end{document}